\RequirePackage{fix-cm}
\documentclass{svjour3}                    
\smartqed  

\usepackage{cite}
\usepackage{amsmath,amssymb,amsfonts}
\usepackage{algorithmic}
\usepackage{graphicx}
\usepackage{adjustbox}
\usepackage{textcomp}
\usepackage{subfigure}
\usepackage{setspace}
\usepackage[figuresright]{rotating}
\usepackage{enumitem}
\usepackage{algorithm}
\usepackage{multirow}
\usepackage{textcomp}
\usepackage[flushleft]{threeparttable}
\usepackage{booktabs,tabularx,tabu}
\usepackage{gensymb}
\usepackage{fixltx2e}
\usepackage{dblfloatfix}
\usepackage{color}
\usepackage{kotex}
\usepackage{caption}

\def\BibTeX{{\rm B\kern-.05em{\sc i\kern-.025em b}\kern-.08em
    T\kern-.1667em\lower.7ex\hbox{E}\kern-.125emX}}
\usepackage{multirow, dcolumn}

\makeatletter
\newcommand*{\rom}[1]{\expandafter\@slowromancap\romannumeral #1@}
\renewcommand{\arraystretch}{1.2}
\makeatother

\makeatletter
\DeclareRobustCommand{\textsupsub}[2]{{%
  \m@th\ensuremath{%
    ^{\mbox{\fontsize\sf@size\z@#1}}%
    _{\mbox{\fontsize\sf@size\z@#2}}%
  }%
}}
\makeatother

\begin{document}

\title{Frame-rate Up-conversion Detection Based on Convolutional Neural Network for Learning Spatiotemporal Features}

\author{Minseok Yoon\textsuperscript{1} \and Seung-Hun Nam\textsuperscript{*,2} \and In-Jae Yu\textsuperscript{1} \and Wonhyuk Ahn\textsuperscript{2} \and Myung-Joon Kwon\textsuperscript{3} \and Heung-Kyu Lee\textsuperscript{*,1}}

\institute{ 
            Minseok Yoon \at
            minseokyoon.8399@gmail.com 
            \and
            Seung-Hun Nam \at
            shnam1520@gmail.com
            \and
            In-Jae Yu \at
            myhome98304@gmail.com
            \and
            Wonhyuk Ahn \at
            whahnize@gmail.com
            \and
            Myung-Joon Kwon \at
            mjkwon2021@gmail.com
            \and
            Heung-Kyu Lee \at
            heunglee@kaist.ac.kr
            \and
            \textsuperscript{*} Seung-Hun Nam and Heung-Kyu Lee are co‐corresponding authors\at
            \and
            \textsuperscript{1} School of Computing, Korea Advanced Institute of Science and Technology (KAIST), Daejeon, South Korea\at
            \and
            \textsuperscript{2} NAVER WEBTOON Corp., Seongnam, South Korea\at
            \and
            \textsuperscript{3} School of Electrical Engineering, Korea Advanced Institute of Science and Technology (KAIST), Daejeon, South Korea
            }
\date{Received: date / Accepted: date}

\maketitle

\begin{abstract}
With the advance in user-friendly and powerful video editing tools, anyone can easily manipulate videos without leaving prominent visual traces. 
Frame-rate up-conversion (FRUC), a representative temporal-domain operation, increases the motion continuity of videos with a lower frame-rate and is used by malicious counterfeiters in video tampering such as generating fake frame-rate video without improving the quality or mixing temporally spliced videos.
FRUC is based on frame interpolation schemes and subtle artifacts that remain in interpolated frames are often difficult to distinguish. Hence, detecting such forgery traces is a critical issue in video forensics.
This paper proposes a frame-rate conversion detection network (FCDNet) that learns forensic features caused by FRUC in an end-to-end fashion.
The proposed network uses a stack of consecutive frames as the input and effectively learns interpolation artifacts using network blocks to learn spatiotemporal features.
This study is the first attempt to apply a neural network to the detection of FRUC. Moreover, it can cover the following three types of frame interpolation schemes: nearest neighbor interpolation, bilinear interpolation, and motion-compensated interpolation.
In contrast to existing methods that exploit all frames to verify integrity, the proposed approach achieves a high detection speed because it observes only six frames to test its authenticity.
Extensive experiments were conducted with conventional forensic methods and neural networks for video forensic tasks to validate our research. The proposed network achieved state-of-the-art performance in terms of detecting the interpolated artifacts of FRUC. The experimental results also demonstrate that our trained model is robust for an unseen dataset, unlearned frame-rate, and unlearned quality factor. Furthermore, FCDNet can precisely localize the tampered region applied to manipulation along the time-domain through temporal localization.

\keywords{Video forensics\and Frame-rate conversion detection\and Frame interpolation scheme\and Convolutional neural network (CNN)\and Residual features\and Spatiotemporal features}
\end{abstract}

\section{Introduction}
\label{sec_introduction}
With the significant spread of video recording devices and video-sharing platforms such as YouTube, TikTok, and Facebook, the number of shared and consumed videos is growing rapidly.
Simultaneously, as editing tools like Adobe Premiere Pro and Final Cut Pro become more popular among the public, even nonprofessionals can easily change the content of videos using editing applications \cite{stamm2013forensic}.
These video-related advances have benefited modern people's lives. However, they are now causing social problems due to contents modified with malicious intent \cite{verdoliva2020media}.
In other words, a manipulated video, whose authenticity is not easily identified by the human eye, can be distributed to cause confusion in fields such as courts, the media, and society \cite{henet,nam2020deep}.
In particular, as deep learning-based tampering techniques such as deepfake have been developed recently \cite{faceforensics,cozzolino_forensictransfer_2019,afchar2018mesonet}, it is widely believed that the abuse cases of intentionally forged video will increase exponentially.
In this situation, most people regard malicious video forgery as a national security concern. Consequently, classifying the manipulation artifacts in forged videos has become a critical topic in multimedia security \cite{verdoliva2020media}.

Therefore, the necessity of a technique for verifying a given video's integrity and authenticity (e.g., whether video manipulation has been applied) is being emphasized \cite{sitara2016digital}.
The related research field is video forensics, which aims to detect and capture forensic clues of various manipulations applied in the spatial and temporal domains of a given video \cite{nam2019video}.
The primary assumption of video forensics is that the original data has inherent characteristics caused by the media acquisition process (e.g., sensor pattern noise \cite{lukas2006digital}, traces of the color filter array \cite{popescu2005exposing}, and compression artifacts \cite{verma_block-level_2020,wang_double_2016}).
These statistical properties are commonly maintained within the media data if no forgery occurs and can be altered when the tampering process is applied.
Therefore, video forensics must capture these subtle clues by focusing on the changes in the statistical patterns of the media \cite{farid2005forensic}.
For a well-behaved video forensic system, it is necessary to study the intrinsic patterns within the media acquisition process while analyzing the characteristics of manipulation artifacts \cite{stamm2013forensic,verdoliva2020media}.
In the past, achieving this goal required conducting the forensic task using handcrafted features specialized in statistical fingerprint extraction and analyzing the periodic patterns retained in the manipulated video \cite{sitara2016digital}.

With a novel perspective, deep learning-based approaches with a convolutional neural network (CNN), which automatically learns forensic features, have been actively studied.
In the early stages of the study, many researchers analyzed of neural network-based methods for detecting image forgery. And the scope of the research varies as follows: image manipulation classification \cite{R2,R3,R5,mayer2019forensic,yu2020manipulation}, double JPEG detection \cite{barni2017aligned,park_double_2018,ahn2019doublejpeg}, computer graphics detection\cite{R7,R8}, content-aware retargeting detection \cite{nam2019seam,nam2020deep}, and camera model identification \cite{R11,R12}.
Compared with conventional methods, the network-based approaches have achieved excellent performance; hence, their interest within CNN-based forensics is steadily increasing.
Recently, as social issues caused by video forgery have become more frequent, image forensic methods have been extended to video forensics.

\begin{figure}[t]
\centerline{\includegraphics[width=0.7\linewidth]{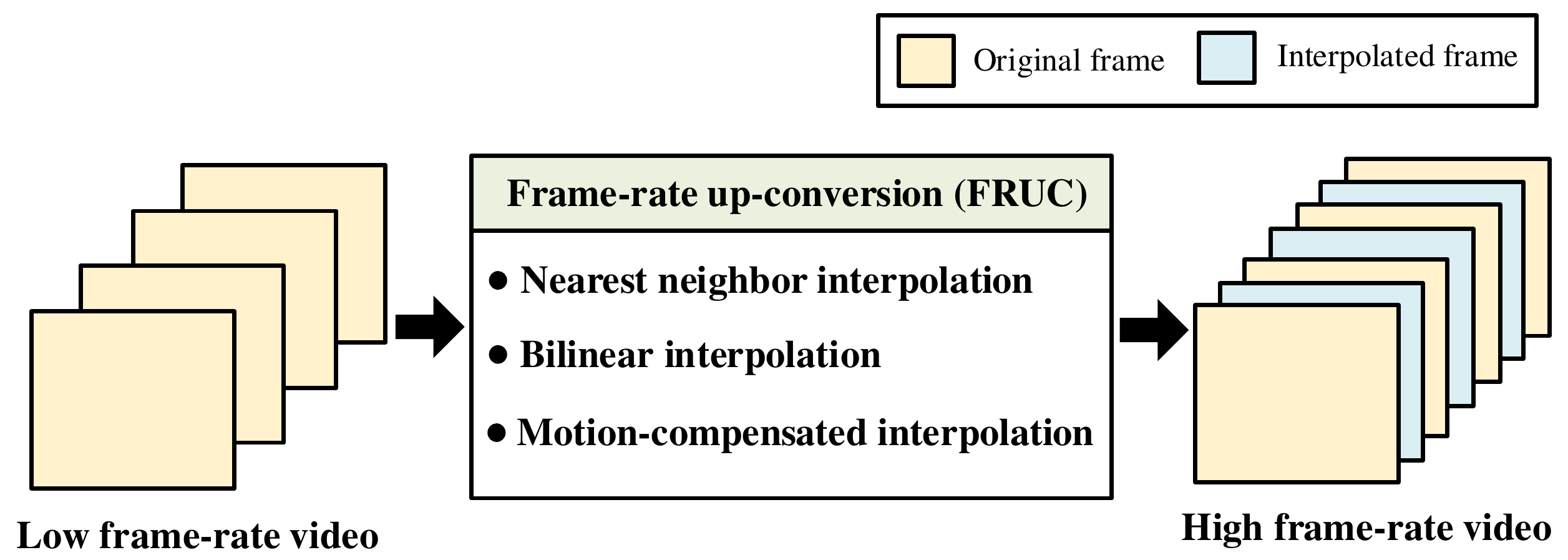}}
\caption{Illustration of FRUC based on frame interpolation schemes. The interpolated frames, generated based on the original frame, are added to the video, making it possible to increase the fps.}
\label{figure_fruc_concept}
\end{figure}

Video forensics is a technique that detects forgery artifacts applied to given videos, with various targets as follows: double compression, frame-rate conversion, inter-frame forgery (e.g., frame insertion or frame deletion), and region tampering \cite{sitara2016digital}.
Frame-rate conversion, a technique to control the frames per second (fps), is a representative temporal domain operation \cite{dar2015motion,yoo2013direction} in video manipulation.
With the advance in high-end display devices and network-related technologies, the sharing of videos with a high frame-rate is actively spreading \cite{huang2008multistage}; hence, frame-rate up-conversion (FRUC) is dynamically used by increasing the motion continuity of videos with a lower frame-rate (Fig.~\ref{figure_fruc_concept}).
However, one abuse case of FRUC is when malicious uploaders increase a lower frame-rate video to a higher one without improving the quality to generate more advertising revenue. Furthermore, FRUC can be used for video manipulation, such as splicing two videos with different frame-rates by up-converting the lower frame-rate video to match the higher ones.
Currently, deep learning-based approaches for detecting recompression \cite{nam2019video,henet}, frame dropping \cite{frame_dropping}, and deepfake \cite{cozzolino_forensictransfer_2019,faceforensics} are presented, while a CNN-based forensic system for detecting FRUC forensic clues does not exist.
Although conventional approaches \cite{Wang,Bian,yao2016detecting,xia2017detecting,ding2018robust,li2018noise,Bestagini,Jung} have exhibited high performance, they did not consider the robustness of the various types of interpolation methods (e.g., nearest neighbor interpolation (NNI), bilinear interpolation (BI), and motion-compensated interpolation (MCI)), which is an important requirement for FRUC detection.
Furthermore, the approaches require considerable processing time to perform detection due to the need to observe all frames.

We overcome these issues by designing, end-to-end, a frame-rate conversion detection network (FCDNet) that can handle various types of interpolation methods simultaneously with a very fast detection speed.
Based on the analysis results for forensic clues of temporal interpolation schemes, we focused on the subtle traces remaining in the relationship of consecutive frames by extending the interpolation artifact analysis in the spatial-domain to the temporal-domain.
Thus, in this study, a stack of multiple consecutive frames is provided as an input of the network.
For effectively learning interpolated artifacts from given data, we design a network architecture specialized for spatial and temporal feature learning. 
The network comprises four types of network blocks to provide fundamental capabilities including residual feature learning, spatiotemporal feature learning, higher feature learning, and classification.
Moreover, a majority voting ensemble is proposed to thoroughly test the forensic evidence, remaining in multiple stacks, for improving detection performance.
Compared with previous studies, the proposed FCDNet achieved state-of-the-art performance for FRUC detection accuracy despite requiring less computation time.
The main contributions are summarized as follows:

\begin{enumerate}[leftmargin=+.3in]
\item The proposed FCDNet is an end-to-end learnable FRUC detection technique which has a state-of-the-art performance based on an efficient network for learning spatiotemporal feature.
It is the first case of a neural network application for FRUC detection.
\vspace{2mm}
\item We propose a FRUC detection scheme that is achieving both high detection speed and high classification performance because FCDNet can verify integrity accurately with a little part of the video. 
\vspace{2mm}
\item FCDNet is robust for various unseen cases (e.g., frame-rate and encoding factors of video quality) that are unlearned parameters in the training stage. 
It is a significant advantage of practicality and availability in the multimedia forensic area. 
\end{enumerate}

The remainder of this paper is organized as follows.
Section~\ref{sec_related_work} reviews relevant previous studies on detecting FRUC.
The analysis of properties of frame interpolation schemes is performed in Section~\ref{sec_analysis_FRUC}, and the proposed forensic framework is presented in Section~\ref{sec_proposed_method}.
In section~\ref{sec_experiments}, the performance of the proposed network is demonstrated by extensive experiments.
Finally, the conclusion and future work are presented in Section~\ref{sec_conclusion}.

\section{Related Work}
\label{sec_related_work}

\begin{figure*}[t]%
\centering%
\subfigure[]{%
  \includegraphics[width=0.31\linewidth]{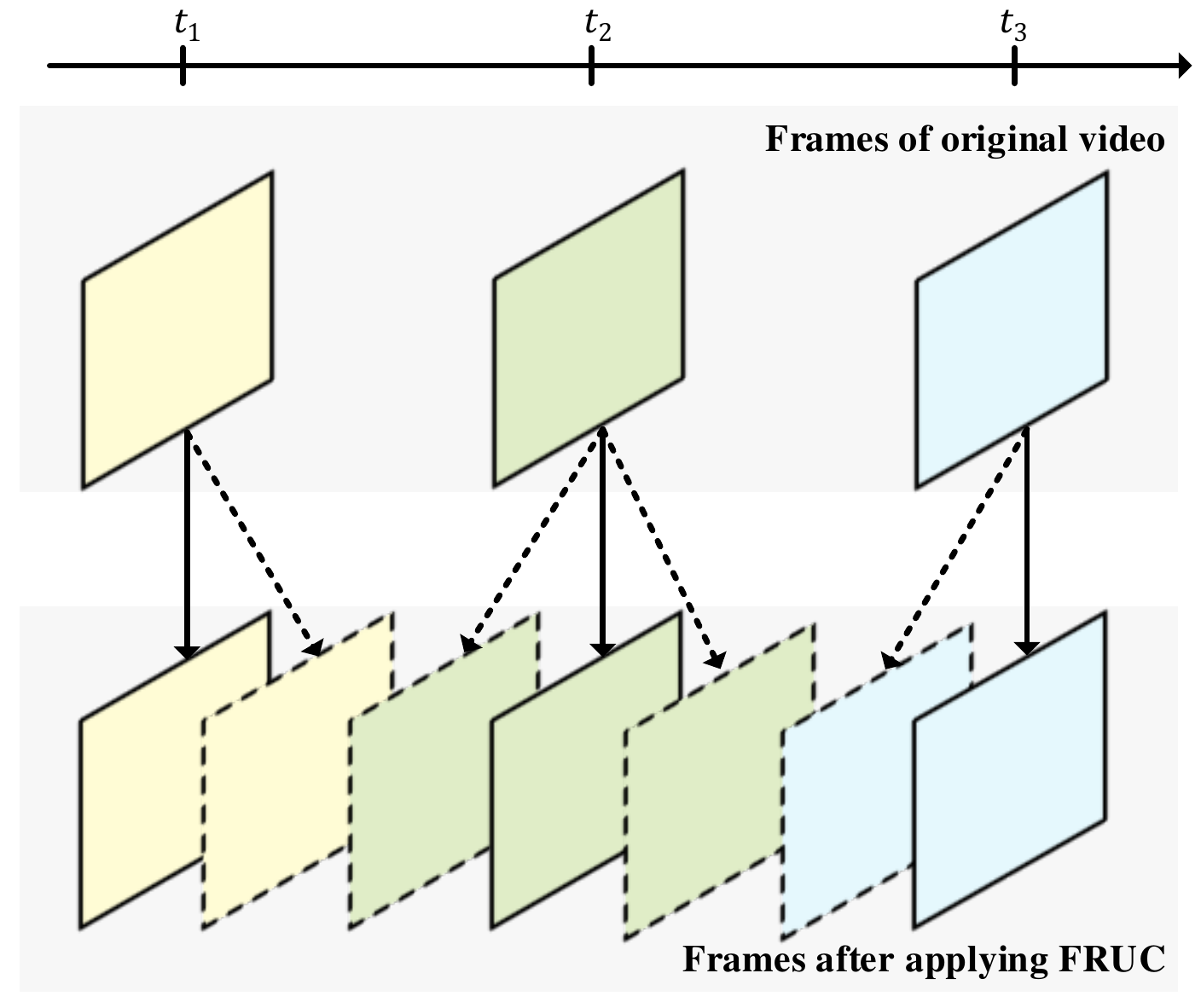}
  \label{fig_frame_interpolation_NNI}
}
\subfigure[]{%
  \includegraphics[width=0.31\linewidth]{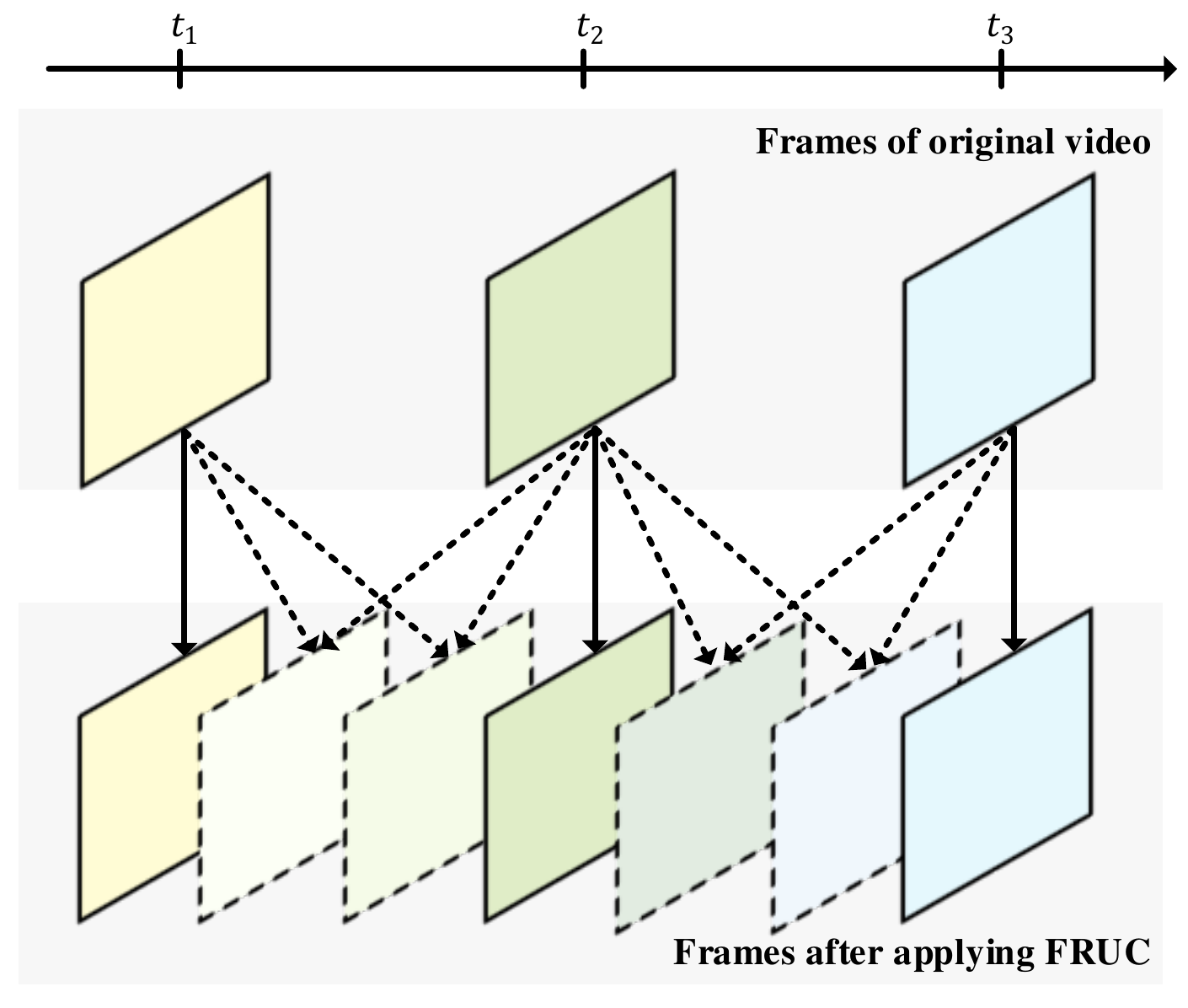}
  \label{fig_frame_interpolation_BI}
}
\subfigure[]{%
  \includegraphics[width=0.31\linewidth]{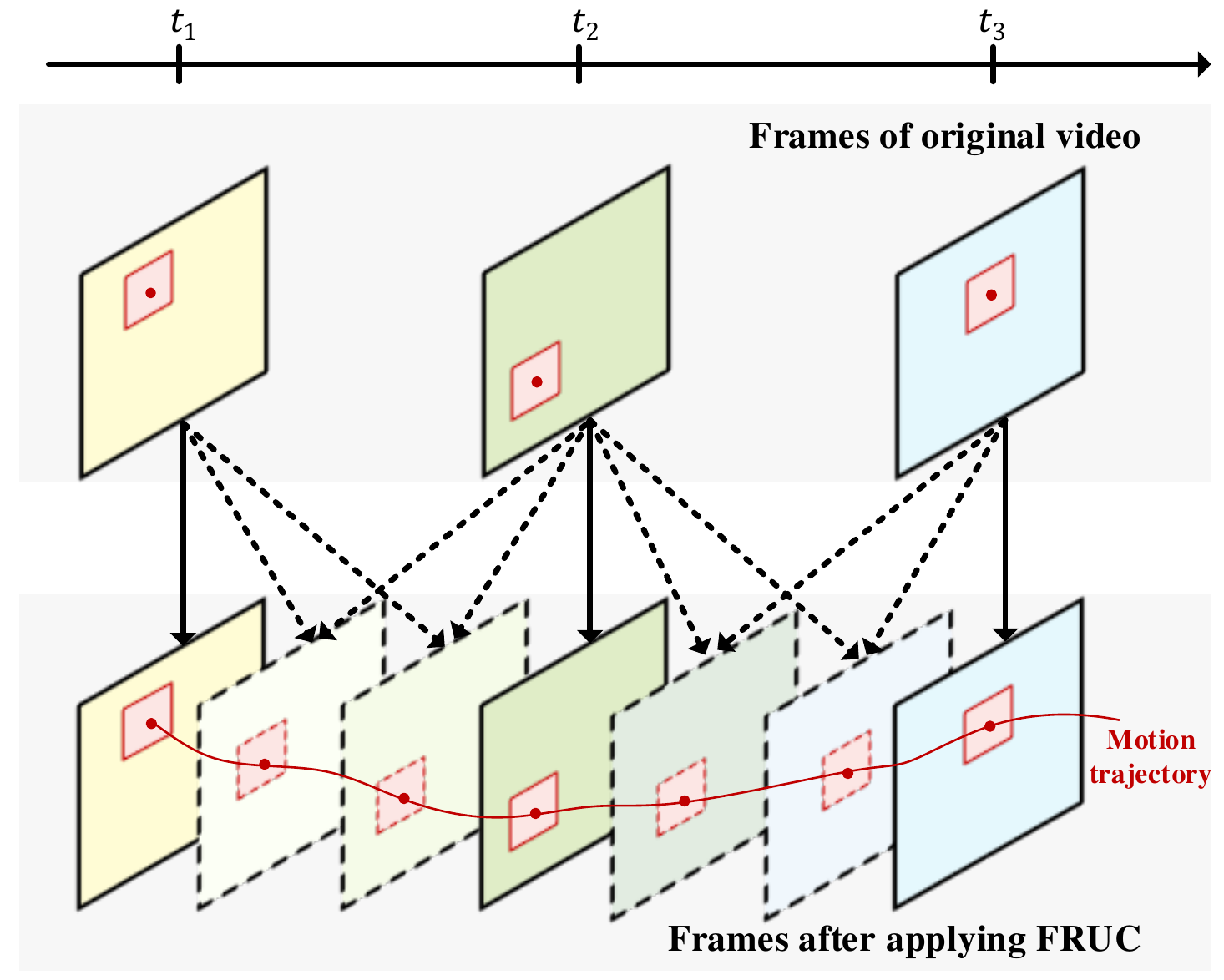}
  \label{fig_frame_interpolation_MCI}
}
\caption{Illustration of frame interpolation schemes used for FRUC: (a) NNI, (b) BI, and (c) MCI. For each scheme, frames indicated by dashed lines represent the interpolated frames.}
\label{fig_frame_interpolation}%
\end{figure*}

Frame-rate conversion, a representative temporal-domain operation, is a promising technique to control the fps. Recently, with the development of display devices and network-related technologies, the distribution and sharing of videos with high frame-rate have been steadily increasing.
Introduced in Section~\ref{sec_introduction}, FRUC refers to increasing the fps to improve visual quality. In particular, it is used to increase the motion continuity of videos with lower frame-rate and also used for video tampering, such as faking a high bit rate and temporal splicing, by malicious counterfeiters.
Frame-rate conversion generates traceable periodic artifacts on the motion trajectories caused by temporal interpolation schemes such as NNI, BI, and MCI.
As depicted in Fig.~\ref{fig_frame_interpolation}, the process of generating and predicting interpolated frames for each method differs. Accordingly, unique manipulation traces remain in the frame-rate converted video.
Furthermore, because frame-rate conversion can be exploited in several video forgery scenarios, detecting and exploring temporal interpolation artifacts is an important issue in the video forensic field.

In the last decade, several blind detection approaches that exploit periodic analysis instead of video-wise or frame-wise analysis have been proposed to identify frame-rate conversion traces. Each technique has advantages and limitations.
Wang \textit{et al.} \cite{Wang} analyzed the motion ratio between the inter-filed motion and the inter-frame motion for detecting frame-rate converted video. However, this approach can only apply to interlaced video.
In addition, the authors presented an expectation-maximization algorithm-based approach to model the relationship between two adjacent frames.
It can identify the traces of FRUC remaining in both interlaced and progressive videos, but it only considers NNI and BI schemes.
Bian \textit{et al.} \cite{Bian} proposed a forensic method that targets interpolated artifacts of NNI based on periodic inter-frame similarity.
The similarity between neighboring frames is measured with a structural similarity index measurement, and periodicity is analyzed in the frequency domain representation with a Fourier transform.

In contrast to the method \cite{Wang,Bian} in which the inter-frame similarity analysis is used for frame averaging and repetition, other approaches based on frame-level analysis have been proposed.
Yao \textit{et al.} introduced a scheme based on the periodic property of edge-intensity or edge discontinuity to determine the interpolated frame's location using a frame-by-frame analysis of FRUC \cite{yao2016detecting}.
Xia \textit{et al.} \cite{xia2017detecting} proposed a method for frame-level investigation using average texture variation for localizing the newly generated frames by temporal interpolation.
Although these two approaches \cite{yao2016detecting,xia2017detecting} perform well for high-quality video, their performance degrades for low-quality or low-motion videos.
Subsequently, Ding \textit{et al.} analyzed residual energy distribution within interpolated frames and modeled temporal inconsistencies in artifact regions using Tchebichef moments as shape descriptors \cite{ding2018robust}.
This method is robust for signal processing operations such as blurring and noise addition. However, the researchers did not consider various types of frame interpolation schemes.

Frame-rate conversion detection methods based on video-level artifacts (e.g., noise-level variation \cite{li2018noise}, prediction error \cite{Bestagini}, and motion artifact \cite{Jung}) also exist.
Based on an analysis of MCI, Li \textit{et al.} proposed a forensic method using noise-level estimation \cite{li2018noise}.
The authors used the periodicity of noise-level variation along the temporal dimension and performed spectrum analysis for extracting the salient spikes.
Bestagini \textit{et al.} \cite{Bestagini} introduced an approach based on prediction error and analyzed the periodicity of motion errors extracted from consecutive frames. Converting the prediction error to the frequency domain can help determine whether the frame-rate conversion occurred. 
Jung \textit{et al.} \cite{Jung} proposed a forensic method using the periodicity of motion artifacts, which considers various interpolation schemes.
First, the motion artifact is computed from the motion vectors of each frame using motion pruning process, and then the periodicity of motion artifacts is analyzed by Fourier transform.
The method demonstrated outstanding performance but is not robust to the BI and requires a large number of frames to detect frame-rate conversion.

The described previous studies \cite{Wang,Bian,yao2016detecting,xia2017detecting,ding2018robust,li2018noise,Bestagini,Jung} have exhibited high performance for specific targeted interpolation algorithms. However, they do not fully satisfy the fundamental requirements of video forensics for detecting frame-rate conversion because they are not robust against various types of interpolation schemes (e.g., NNI, BI, and MCI).
Furthermore, the approaches require considerable processing time to perform detection because they can achieve proper performance only after all frames of a video are analyzed.
To address the drawbacks of the conventional approaches using predefined rules and handcrafted features, we propose a CNN-based forensic approach to enable the network to learn interpolation artifacts automatically.
The proposed FCDNet, consisting of network blocks for learning spatiotemporal features, can effectively explore temporal interpolation artifacts left in interpolated frames and cover three types of interpolation methods specified in Fig.~\ref{fig_frame_interpolation}.
Moreover, because FRUC detection is performed by requiring only a few frames, it has a higher computational speed than existing methods.

\section{Analysis of Interpolation Artifacts}
\label{sec_analysis_FRUC}
In this section, we review frame interpolation schemes used in FRUC, including NNI, BI, and MCI. We also analyze forensics clues remaining in interpolated frames. Based on the analysis results, we suggest an efficient approach to design a network architecture and learn spatiotemporal features.

\begin{figure*}[t]
\centerline{\includegraphics[width=\linewidth]{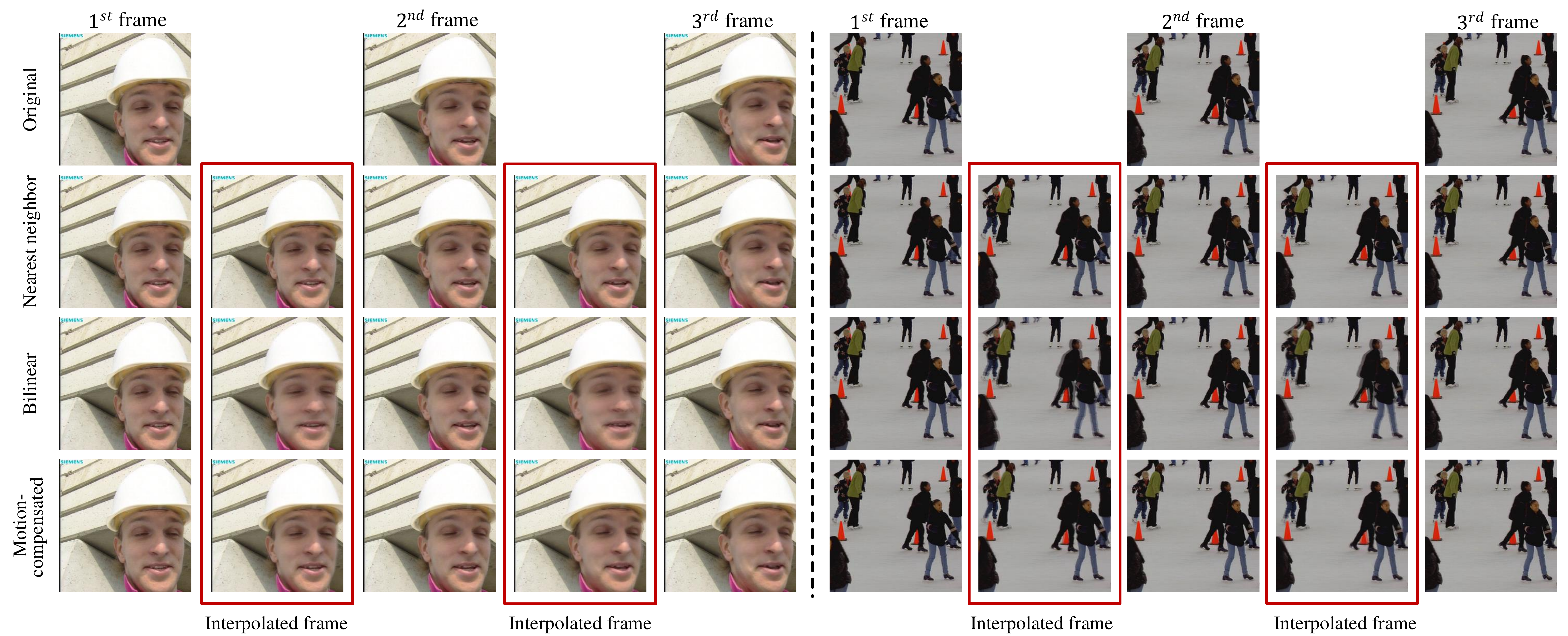}}
\caption{Example of interpolated frames generated based on frame interpolation techniques NNI, BI, and MCI. Interpolated frames (red boxes) using different types of interpolation methods leave a trace of the synthesized frames.}
\label{fig_interpolated_frame}
\end{figure*}

\subsection{Review of Frame Interpolation Scheme} 
\label{sec_interpolation_scheme}
The FRUC is performed with frame interpolation schemes between adjacent video frames to increase the motion continuity of low frame-rate video and improve the visual quality on displays. Moreover, compression standards for video such as MPEG-4 part 2 and H.264, are built into these interpolation schemes.
The goal of frame interpolation is to produce interpolated frames in the middle of two adjacent frames of the original video while minimizing visual artifacts.
The proposed scheme aims to detect FRUC by observing interpolation artifacts remaining in such interpolated frames.
Accordingly, it is essential to analyze the characteristics and the operation process of each interpolation scheme.

In this section, descriptions of frame interpolation schemes (i.e., NNI, BI, and MCI) are provided.
The selected interpolation schemes are applied in various compression standards and video editing tools.
As depicted in Fig.~\ref{fig_frame_interpolation}, each scheme has different operations and complexities for generating interpolated frames.
Thus, the visual characteristics of the interpolated frames generated by each interpolation scheme differ considerably (Fig.~\ref{fig_interpolated_frame}).
For analyzing the forensic clues caused by frame interpolation, a brief review of each scheme is presented.

\subsubsection{Nearest Neighbor Interpolation}
\label{sec_nni}
NNI is the simplest frame interpolation scheme.
As illustrated in Fig.~\ref{fig_frame_interpolation_NNI}, this scheme duplicates the closest frame in a temporal direction among the original frames and then places the duplicated frame at the intermediate frame position.
Rather than calculating an average value by some weighting criteria or synthesizing an intermediate value based on complicated rules, this scheme uses the nearest neighboring frame as the interpolated frame.
In other words, this approach duplicates and rearranges original frames without synthesizing the new frames.
Thus, it has the lowest computational complexity among the interpolation schemes covered in this section. The interpolated frame quality is excellent in the spatial-domain (Fig.~\ref{fig_interpolated_frame}).
However, motion discontinuity increases in the time-domain.

\subsubsection{Bilinear Interpolation}
BI synthesizes the interpolated frame by linearly combining both sides of the original frames (i.e., the previous frame and the next frame) in the temporal direction (Fig.~\ref{fig_frame_interpolation_BI}).
The estimated flows-based warping operations with BI are performed first. Then, the warped frames are blended to generate the interpolated frame.
This interpolation scheme enables a more natural scene flow than NNI and is widely used because of its simplicity and low computational complexity.
However, as depicted in Fig.~\ref{fig_interpolated_frame}, it has the disadvantage that the interpolated frames are degraded by ghost artifacts and blurry artifacts.
These blurry artifacts predominantly occur when warped frames are not well aligned due to errors in the estimated flows.

\subsubsection{Motion-Compensated Interpolation}
MCI is a technique for generating the intermediate frame $f_n$ between the previous frame $f_{n-1}$ and the next frame $f_{n+1}$ based on a motion vector (Fig.~\ref{fig_frame_interpolation_MCI}).
To do this, several fundamental elements are required: motion estimation (ME), motion vector smoothing (MVS), and MCI \cite{yoo2013direction,huang2008multistage}.
As depicted in \cite{dar2015motion}, ME based on a block-matching algorithm is used to estimate motion vectors between adjacent frames (e.g., $f_{n-1}$ and $f_{n+1}$) to measure object motion. MVS is used to refine the computed motion vectors along the spatiotemporal direction \cite{li2018noise}.
MCI then synthesizes the intermediate frame $f_n$ using refined motion vectors. This process is as follows: $f_{n}(x,y)=\alpha\cdot f_{n-1}(x+u_{x},y+u_{y})+\beta\cdot f_{n+1}(x+v_{x},y+v_{y}),$ where $(u_{x},u_{y})$ and $(v_{x},v_{y})$ indicate the motion vectors pointing to the $f_{n-1}$ and $f_{n+1}$ in the horizontal and vertical directions, respectively.
$\alpha$ and $\beta$ are weighting factors whose sum is equal to one.
As depicted in Fig.~\ref{fig_interpolated_frame}, this scheme can generate more natural frames than BI but requires high computational complexity.

\begin{figure*}[t!]%
\centering%
\subfigure[Nearest neighbor interpolation]{%
  \includegraphics[height=1.8in]{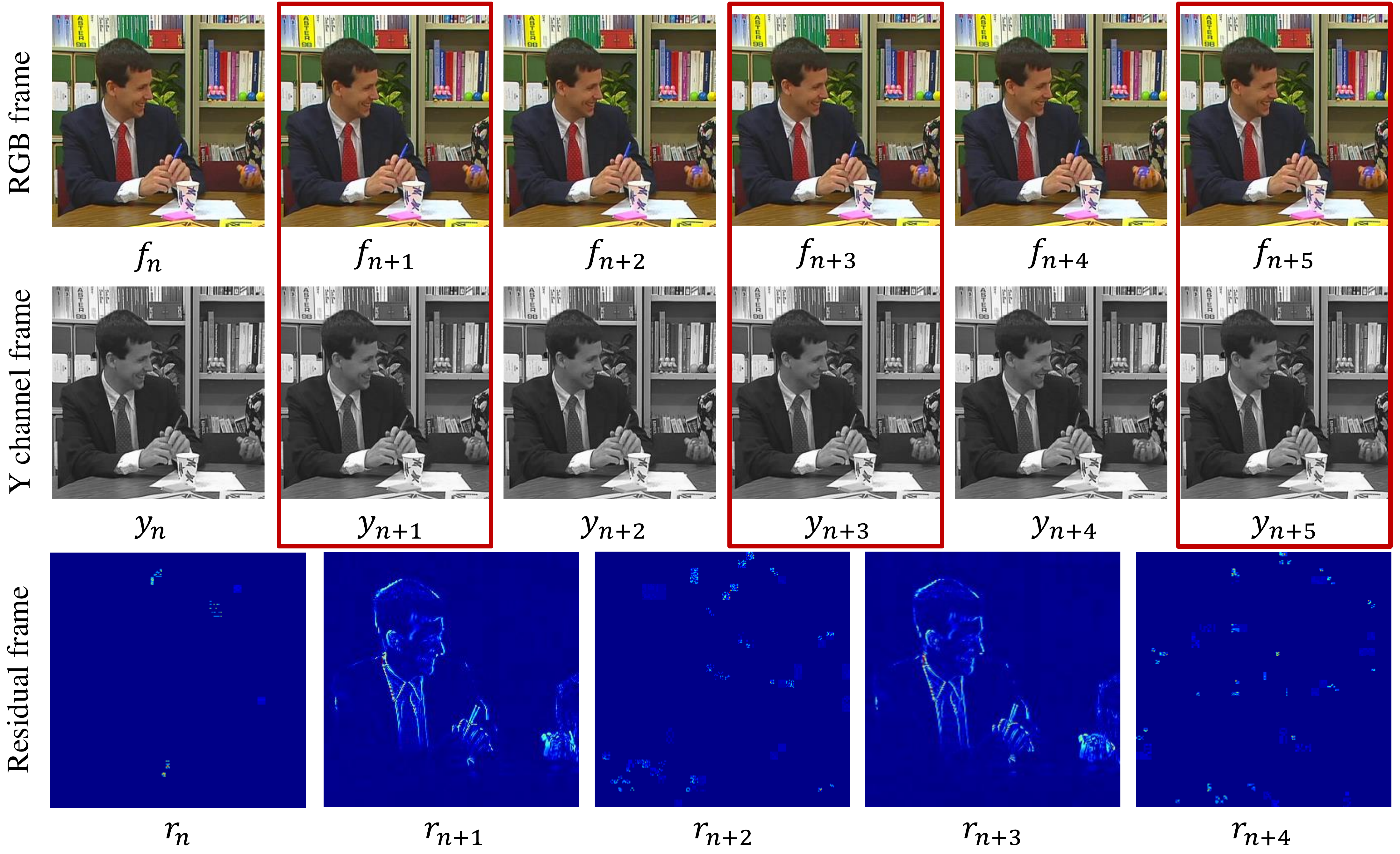}
  \label{fig_frame_residual_data_NNI}
}
\subfigure[Bilinear interpolation]{%
  \includegraphics[height=1.8in]{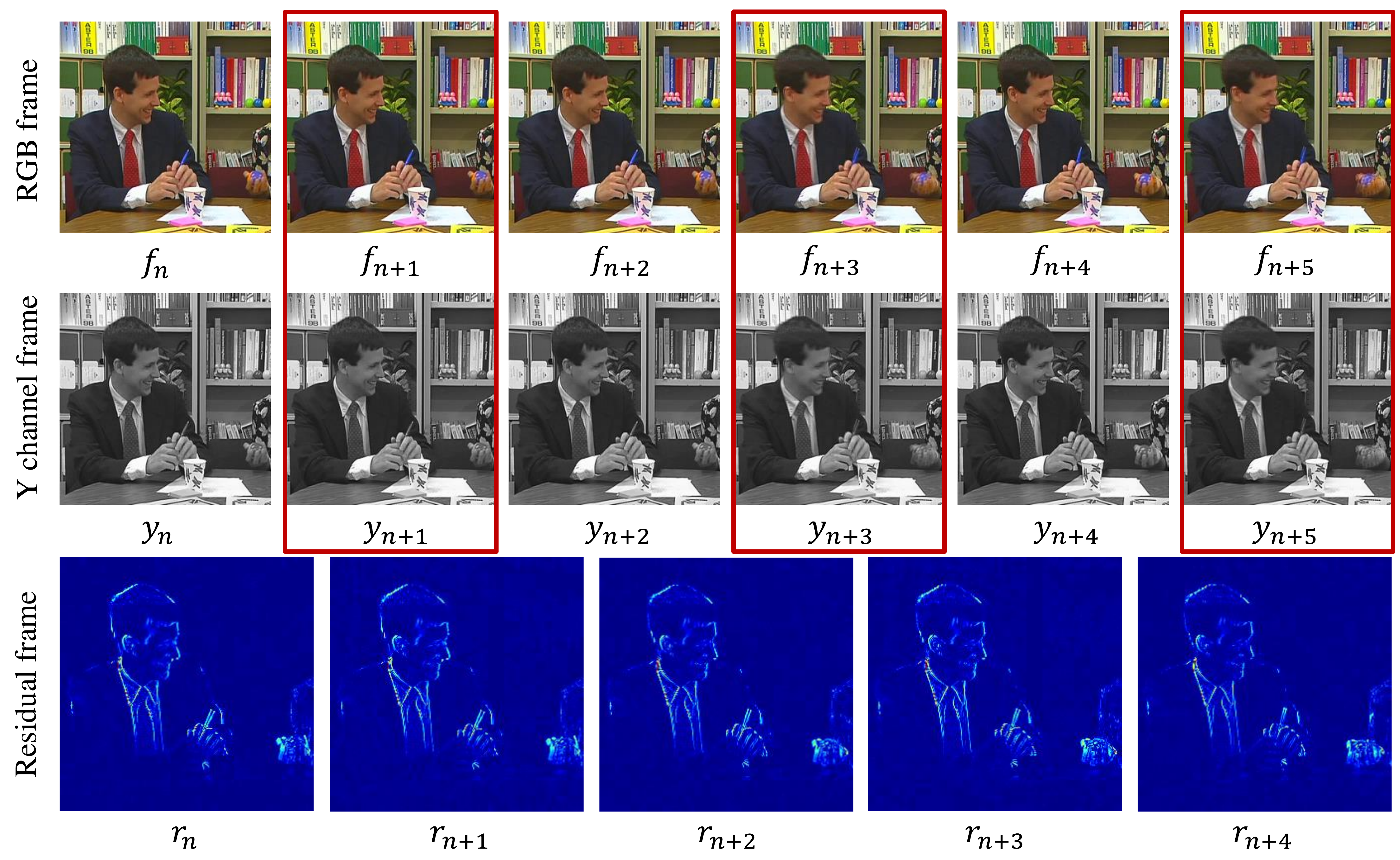}
  \label{fig_frame_residual_data_BI}
}
\subfigure[Motion-compensated interpolation]{%
  \includegraphics[height=1.8in]{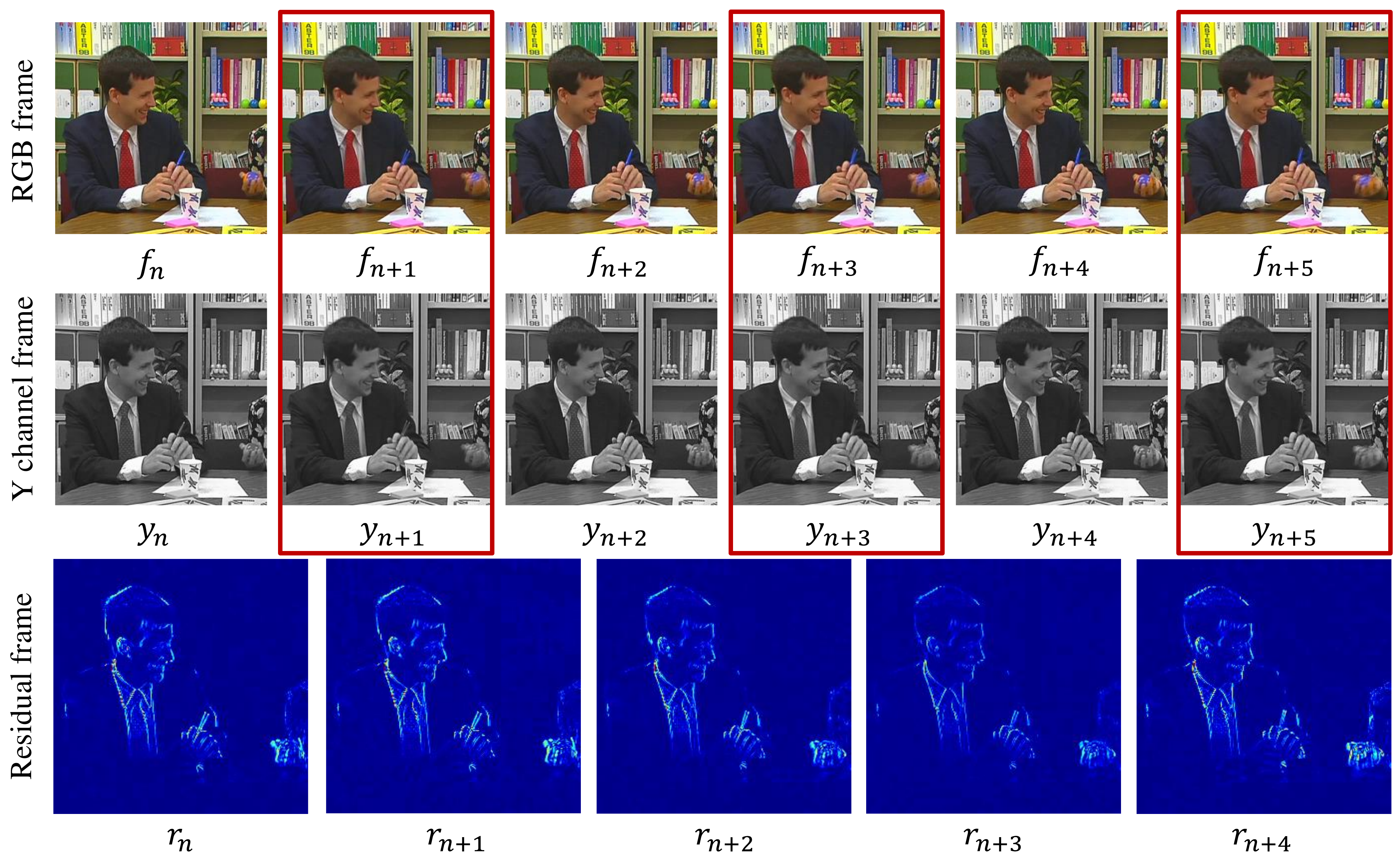}
  \label{fig_frame_residual_data_MCI}
}
\caption{Examples of residual frames generated from three types of interpolation schemes: (a) NNI, (b) BI, and (c) MCI. From the top to bottom rows, each result represents RGB frame, Y-channel frame, and residual frame, respectively. The interpolated RGB frames and Y-channel frames are in red boxes. The residual frame $r_{n}$ is calculated by $r_{n} = y_{n} - y_{n+1}$; $n$ is the index of the frame.}
\vspace{-3mm}
\label{figure_residual_data}%
\end{figure*}

\subsection{Forensics Clue of Interpolated Frame}
For detecting the traces of FRUC using the interpolation schemes described in Section~\ref{sec_interpolation_scheme}, we analyzed the properties of synthesized frames generated from each scheme.
As depicted in the red boxes of Fig.~\ref{fig_interpolated_frame}, the synthesized frames' visual characteristics are diverse with each interpolation method.
We compared the interpolation artifact results remaining on the spatial-domain according to three types of interpolation methods. NNI scheme did not leave any visual evidence theoretically because it copied the nearest frame and pasted the frame as an interpolated frame \cite{Jung}.
Furthermore, we observe the ghost and blurry artifacts in the interpolated frame caused by the other frame interpolation schemes, BI and MCI, used in the previous and the next frame to generate the interpolated frame (the 3-$rd$ and 4-$th$ rows in Fig.~\ref{fig_interpolated_frame}).
In particular, these ghost artifacts and blurry artifacts, caused by the combination of the multiple frames with moving objects, were found prominently in the BI method.
Therefore, to detect artifacts caused by various interpolation schemes, an approach is required that effectively learns subtle signals remaining in the spatial-domain.

Recently, with the development of deep learning, CNN approaches have been introduced that capture and learn low-level signals remaining in an image \cite{mislnet,srnet,nam2019seam,ahn2020local} and a single frame of video \cite{henet,nam2019video,afchar2018mesonet}.
The use of these networks can facilitate learning the blurry artifacts in the spatial-domain caused by BI and MCI schemes.
However, as stated in Section~\ref{sec_interpolation_scheme}, there are no visually prominent clues in the interpolated frame obtained from NNI, which merely duplicates the original frame. There exist only subtle recompression artifacts.
Hence, CNN-based approaches for learning low-level features remaining in a single frame cannot be directly applied to the frame-rate converted video used in NNI method.

We addressed this issue by focusing on the forensic clues remaining in the consecutive frames and considered the interpolation artifacts in both spatial-domain and temporal-domain. Compared with the other two approaches considering natural motion continuity, NNI duplicates and rearranges original frames without a synthesis process. Therefore, it has the drawback of motion discontinuity in the temporal-domain. This motion discontinuity is visually observed in residual frames between successive frames of the frame-rate converted video.
Based on the visualization of residual frames as depicted in Fig.~\ref{fig_frame_residual_data_NNI}, an approach to observing the difference between adjacent frames would be beneficial for FRUC detection. Because forensic clues caused by three interpolation schemes exist in both spatial and temporal domains, a stack of multiple consecutive residual frames is provided as an input to the proposed FCDNet. Moreover, to capture various interpolation artifacts from the inputs, we designed a network structure suitable for learning abundant spatiotemporal features. The details are introduced in the following sections.

\begin{figure}[t]
\centerline{\includegraphics[width=0.8\linewidth]{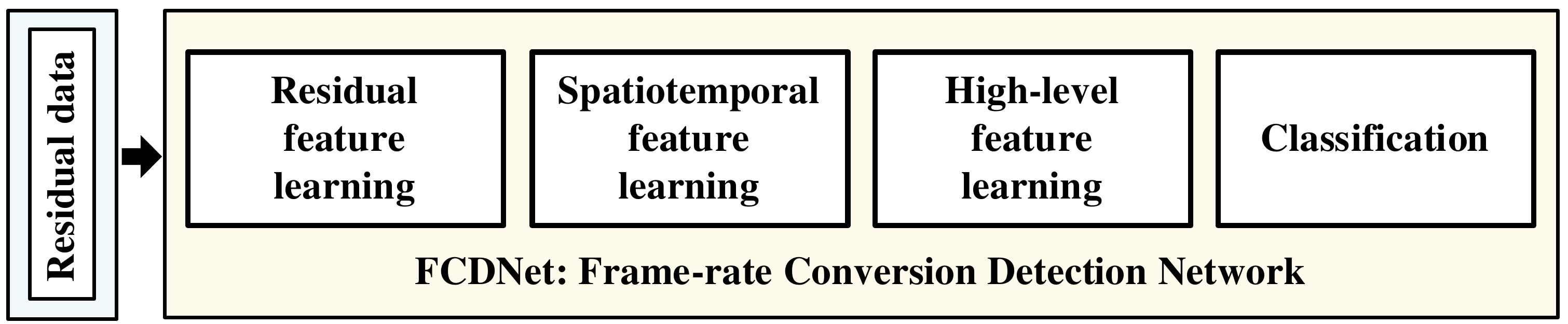}}
\caption{Overview of the proposed FCDNet for detecting FRUC. 
Our model, which uses residual data as an input, comprises network blocks for residual feature learning, spatiotemporal feature learning, higher feature learning, and classification.}
\label{figure_overview}
\end{figure}

\section{Proposed Forensic Framework}
\label{sec_proposed_method}
In this paper, we propose a forensic framework for detecting FRUC artifacts caused by various frame interpolation schemes.
Accordingly, we propose an FCDNet specialized in learning spatiotemporal features that uses multiple frames, successively accumulated residual frames, as an input. Fig.~\ref{figure_overview} illustrates an overview of FCDNet. The network comprises several types of network blocks to enable the capabilities of residual feature learning, spatiotemporal feature learning, high-level feature learning, and classification. Furthermore, a majority voting ensemble is presented to comprehensively analyze the forensic clues remaining in the multiple stacks of input frames through the temporal-domain to improve detection performance.

\subsection{Residual Input Data}

As depicted in Figs.~\ref{fig_interpolated_frame} and \ref{figure_residual_data}, when a frame-rate conversion occurs in video, it is challenging to distinguish between the original and the forged video by analyzing the frame-by-frame due to the development of precise interpolation techniques.
Moreover, it is almost impossible for NNI to classify through a network framework that learns frame-by-frame because an interpolated frame is created by just copying a frame from an original video.
In our work, we design FCDNet that learns the correlation between residual frames to use the forensic clues of interpolated frames analyzed in the previous section. The detailed process of residual input feature generation consists of three steps as follows.

\begin{enumerate}[leftmargin=+.3in]
\item Extract the input frames $I$ composed of the six consecutive frames \{$f_{n}$, $f_{n+1}$, ..., $f_{n+4}$, $f_{n+5}$\} from a test video, where $n$ $\in$ \{1, 2, 3, ..., $N$-5\} and $N$ is the number of frames in the video. The first frame of the consecutive frames ($f_{n}$) is selected from a random position in the video.
We refer to feature maps with a size of ($t$$\times$$c$$\times$$h$$\times$$w$), where $t$ is the number of frames, $c$ is the number of channels in one frame, and $h$ and $w$ are the height and width of the frame, respectively. $f_{n}$ has three RGB channel and height and width are 256 so that the shape of $f_{n}$ and $I$ are (3$\times$256$\times$256) and (6$\times$3$\times$256$\times$256). 

\item RGB frames with three channels are converted into gray frames with one channel. The converted input frames $Y$, \{$y_{n}$, $y_{n+1}$, ... , $y_{n+4}$, $y_{n+5}$\}, is obtained by calculating each luminance frame $y_{n}$ using equation $y_{n}=f_{n}$ $\times$ $[0.299, 0.587, 0.114]$.
The feature map of $Y$ is (6$\times$1$\times$256$\times$256). We then reshape it into (6$\times$256$\times$256) as a dimensionality reduction. Consequently, we can control the 3D video as a 2D image.

\item The residual frame $r_{n}$ typically is calculated by the difference between consequence luminance frames $y_{n}$ and $y_{n+1}$.
The bundle of five consecutive residual frames \{$r_{n}, r_{n+1}, ... , r_{n+4}$\} are used to FCDNet's input $I_{R}$.
The $I_{R}$ includes not only spatial but also temporal information and the shape of $I_{R}$ is (5$\times$256$\times$256).
\end{enumerate}

We improved performance by producing a stack of residual frames that contain motion-specific information by removing objects and backgrounds in the video content. Frame-rate conversion is a representative temporal-domain operation. Consequently, we have to focus more on the temporal information using the difference between frames. The generation of residual frames is equal to the extraction of movement information in a bundle of input frames. The residual frames help to reduce the video dimensions which is an advantage in terms of computation efficiency.
The movement information exists in the spatial axis in a single residual frame, whereas it exists in both spatial and temporal axis in stacked residual frames.
When the same interpolation technique is used in video frames, there is a specific statistical correlation between consecutive residual frames. 
Therefore, FCDNet uses the correlation between frame differences as an input. The first step of FCDNet applies a convolutional layer to extract subtle evidence in each residual frame. The second step of FCDNet merges the residual features extracted in the consecutive residual frames and learns the spatiotemporal features similar to using a convolutional layer with stacked residual frames. Residual frames are significantly more efficient than the other types of frames for the network's input parameter.

\begin{figure*}[t]
\centerline{\includegraphics[width=1.0\linewidth]{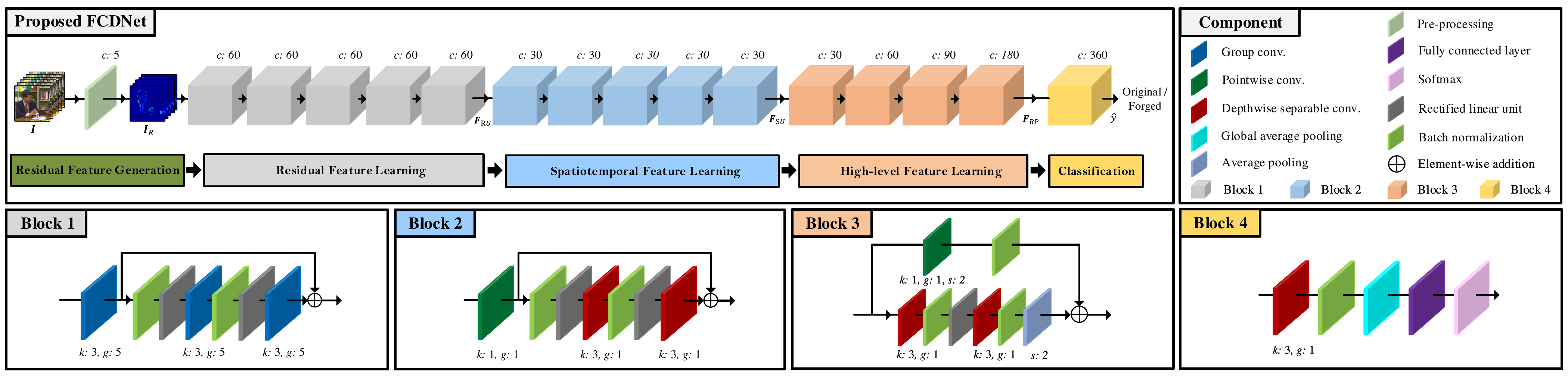}}
\caption{Architecture of the proposed FCDNet for detecting interpolated artifacts caused by frame-rate conversion: $\emph{c}$, $\emph{s}$, $\emph{k}$, and $\emph{g}$ are the number of output feature maps, stride of each layer, kernel size, and the number of groups, respectively.}
\label{figure_architecture}
\end{figure*}

\subsection{Network Architecture}
This section introduces the overall structure of the proposed FCDNet, as depicted in Fig.~\ref{figure_architecture}. 
We propose a FRUC detection method using an efficient spatiotemporal residual framework for learning subtle artifacts generated by the FRUC process in an end-to-end way. FCDNet comprises four parts:

\begin{enumerate}
\item Residual feature learning: Block 1
\item Spatiotemporal feature learning: Block 2
\item High-level feature learning: Block 3
\item Classification: Block 4
\end{enumerate}

We define $I$ as input frames that comprise sequential six frames extracted from one video clip. We add one pre-processing module to generate the residual input frames ($I_{R}$) from consecutive adjacent frames in suspicious input video clips.
\begin{equation}
\label{Eq1}
    I_{R} = H_{RFG} (I), \\
\end{equation}
where $H_{RFG}(\cdot)$ is the residual frame generation module, including a gray scale transformation and a residual frame extraction. The produced five residual input frames $I_{R}$ are used for feature extraction without pooling using the residual feature extraction module $H_{RFE}$ and using the spatiotemporal feature extraction module $H_{SFE}$. 
\begin{equation}
\label{Eq2}
	F_{RU} = H_{RFE} (I_{R}), \\
\end{equation}
\begin{equation}
\label{Eq3}
	F_{SU} = H_{SFE} (F_{RU}), \\
\end{equation}
where $F_{RU}$ and $F_{SU}$ are the extracted unpooled features for residual and spatiotemporal features, respectively.
Unpooling the feature maps contributes to capturing the subtle signal caused by frame interpolation schemes.
$H_{RFE}(\cdot)$ and $H_{SFE}(\cdot)$ consist of one adjustment operation and one residual unit with a skip connection. The adjustment operation is used to control the number of channels at the first stage of Block 1 and 2.

The adjustment operation of $H_{RFE}$ is a 3$\times$3 grouped convolution operation, where the number of groups is set to five. The input channel is the same as the number of groups, the output channel is a multiple of the number of groups, and padding/stride is set to one denoted in Block 1. The grouped convolution operation help to efficiently extract the intra-frame features for each residual input frame.
The residual unit of $H_{RFE}$ consists of two sets of batch normalization (BN), rectified linear unit (ReLU), and 3$\times$3 grouped convolution operation with a stride of one. Moreover, we inject a skip connection between the beginning and end of the residual unit, which is summed with the output of the second convolution operation. 
The skip connection leads to feature reusability, training stability, and enabling convergence.
All feature maps are pre-activated by BN and ReLU before the convolution from the residual unit of Block 1 in Fig.~\ref{figure_architecture}.

The extracted residual features with unpooling $F_{RU}$ are applied to the feature extraction module for spatiotemporal features. $H_{SFE}$ is almost equal to $H_{RFE}$ but differs in convolution type and parameter. 
The adjustment operation of $H_{SFE}$ is a 1$\times$1 point-wise convolution operation with specified parameters depicted in Block 2. The previously mentioned grouped convolution in Block 1 only merges intra-frame information. In contrast, the point-wise convolution decreases the output channels to save on memory and combines inter-channel information which are extracted features from residual input frames. It is used to arrange the number of channels before the residual unit.
The residual unit of $H_{SFE}$ are structurally the same as those of $H_{RFE}$, but a 3$\times$3 depth-wise separable convolution \cite{chollet2017xception} operation are applied to manage spatiotemporal features, in contrast to $H_{RFE}$, which extracts spatial features. The skip connection is an identity mapping function with a stride of one.

The extracted spatiotemporal features with unpooling $F_{SU}$ are then used for feature map reduction modules. 
\begin{equation}
\label{Eq4}
	F_{RP} = H_{FMR} (F_{SU}), \\
\end{equation}
where $H_{FMR}(\cdot)$ and $F_{RP}$ represent the feature map reduction module and the reduced feature with pooling, respectively.
$H_{FMR}(\cdot)$ consists of a 3$\times$3 depth-wise convolution operation for high-level feature learning and a pooling operation for reducing the feature maps.
In Block 3, we scale down the spatial size of feature maps and scale up the channel size of feature maps by adopting convolution operation with a stride of two and averaging pooling operation.

Finally, the reduced and pooled feature $F_{RP}$ is put sequentially through a 1D feature reduction unit $H_{R1D}(\cdot)$ and a linear classification unit $H_{LC}(\cdot)$.
\begin{equation}
\label{Eq5}
	\widehat{y} = H_{LC}(H_{R1D}(F_{RP})), \\
\end{equation}
where $\widehat{y}$ is the probability of binary classification used to verify whether a suspicious video is an original or a forgery.
The 1D feature reduction unit $H_{R1D}(\cdot)$ applies a global averaging pooling (GAP) operation \cite{global} that is more meaningful and interpretable because it enforces an association between feature maps and categories. Furthermore, because GAP is itself a structural regularizer, it natively prevents overfitting for the overall structure.
And, the linear classification unit $H_{LC}(\cdot)$ consists of a softmax function and a fully connected (FC) operation.
The goal of training FCDNet is to minimize the binary cross-entropy loss for a training set of videos and the loss function is optimized by Adam \cite{adam}.
The proposed network FCDNet automatically explores forensic features, end-to-end, using various types of blocks.

\subsubsection{Residual Feature Learning}

The residual feature learning consists of five Block 1's as displayed in Fig.~\ref{figure_architecture}. Block 1 comprises a 3$\times$3 grouped convolution with a stride and padding of one to extract residual features, BN to alleviate the potential of overfitting, and ReLU as an activation function.
We apply the grouped convolution to learn the intra-frame information, in which the number of groups is set to five, that is equal to the number of consecutive residual frames as an input. That is to say, we execute the convolution operation not in the channel direction but only in the spatial direction to extract subtle features within each residual frames.
Block 1 does not use any pooling operation and maintains the spatial size of feature maps (256$\times$256) to avoid removing the fine traces of FRUC.
Block 1 is analogous to Block 2 except for the type of convolution operation but has higher channel information ($c$=60). The high channel capacity in Block 1 can be interpreted as a stronger ability of feature representation in each residual input frame. Therefore, Block 1 has not only a large feature map size but also a high channel resolution through the network hierarchy to achieve excellent accuracy for FCDNet.

\subsubsection{Spatiotemporal Feature Learning}

The spatiotemporal feature learning consists of five Block 2's, as displayed in Fig.~\ref{figure_architecture}. Block 2 is structurally similar to Block 1. Block 2 uses a 1$\times$1 point-wise convolution with a stride/padding of one to learn the association between the feature maps, a 3$\times$3 depth-wise separable convolution with a stride/padding of one to learn the relationship between neighboring elements, a BN to alleviate the potential of overfitting, and a ReLU as an activation function.
We apply the point-wise convolution operation to merge the feature maps only in the channel direction. Because the channel information means the temporal feature of the video in Block 2, we can mix the temporal-wise feature through a point-wise convolution operation.
Moreover, we use the depth-wise separable convolution to learn both intra- and inter-frame feature. In other words, we execute the convolution operation both in the spatial and channel directions.
The depth-wise separable convolution divides the normal convolution process into two parts: a depth-wise convolution and a point-wise convolution. With this technique, we can save computational power and improve efficiency without significantly reducing effectiveness.
Also, Block 2 has a high feature map size (256$\times$256) by excluding the pooling operation to prevent the removal of noise-like interpolation evidence.
In contrast, Block 2 can use a lower channel capacity ($c$=30) than Block 1, that makes the FCDNet lightweight.

\begin{figure}[t]
\centerline{\includegraphics[width=0.8\linewidth]{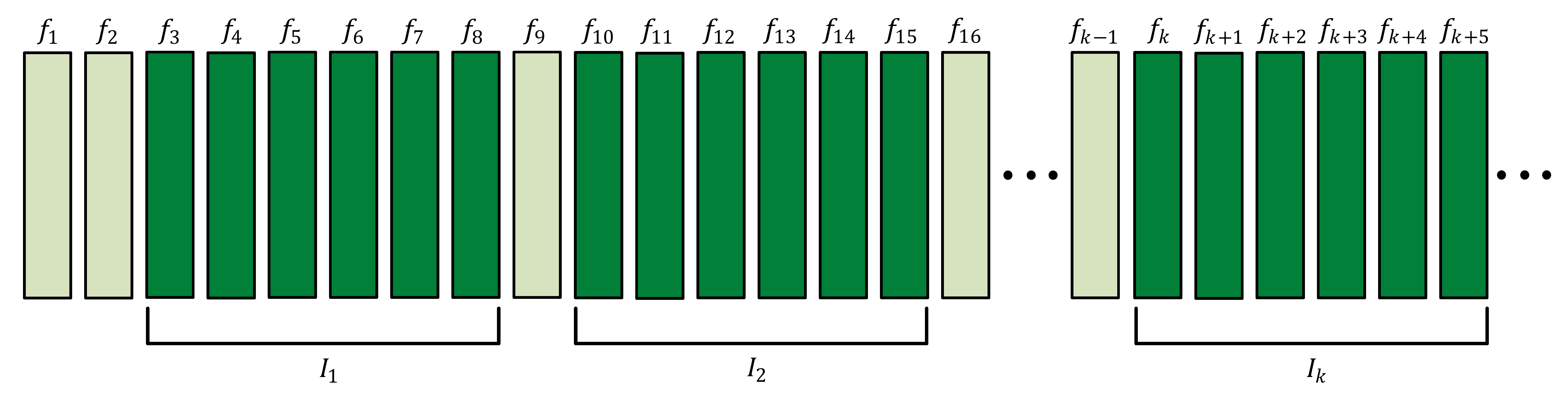}}
\caption{Example of extracting $I_{n}$ in the target video.}
\label{fig_majority_voting_1}
\end{figure}

\subsubsection{High-level Feature Learning}

The high-level feature learning consists of four Block 3's, as depicted in Fig.~\ref{figure_architecture}. 
We use a 3$\times$3 depth-wise separable convolution with a stride/padding of one to learn the high-level feature and two types of pooling operations to reduce the feature map size.
Block 3 is divided into two paths that include different downsampling methods. The first path uses a 1$\times$1 point-wise convolution with a stride of two and a BN operation for concise downsampling to perform element-wise addition. The second path consists of a sequence of 3$\times$3 depth-wise separable convolution, BN, ReLU, and averaging pooling operations to decrease the number of feature maps by downsampling.

\subsubsection{Classification}
The classification consists of one Block 4, as visualized in Fig.~\ref{figure_architecture}. 
We use a 3$\times$3 depth-wise separable convolution with a stride/padding of one to synthesize the high-level features, a GAP operation to alleviate overfitting and improve the generalization ability, a FC operation to merge all extracted features, and a softmax function to transform the output of FC operation to a probability.
Note that the average statistical moments are computed for each channel of extracted feature maps through the GAP in Block 4. Block 1 - 3 have shortcut connections, but Block 4 excludes any shortcut connections.

\subsection{Majority Voting Ensemble}

The proposed FCDNet determines whether the video frame-rate is changed from input frames $I$, which is a stack of six consecutive Y-channel frames. For composing a bundle of five residual frames as the network's input units $I_{R}$, the $I$ is passed the pre-processing module that consists of gray scale transformation step and residual frames extraction step.
The detection accuracy depends on how much frame information is aggregated and utilized. Therefore, the network should extract and use as many $I_{n}$ as possible from one manipulated video to increase the detection accuracy of the FRUC.
However, it has drawbacks; for example, the evaluation speed is slowed down when testing with more frames in the video. Thus, we should choose an appropriate number of $I_{n}$ according to the trade-off between detection time and detection accuracy. In FCDNet, we experimentally set the number of $I_{n}$ to nine.

\begin{figure}[t]
\centerline{\includegraphics[width=0.8\linewidth]{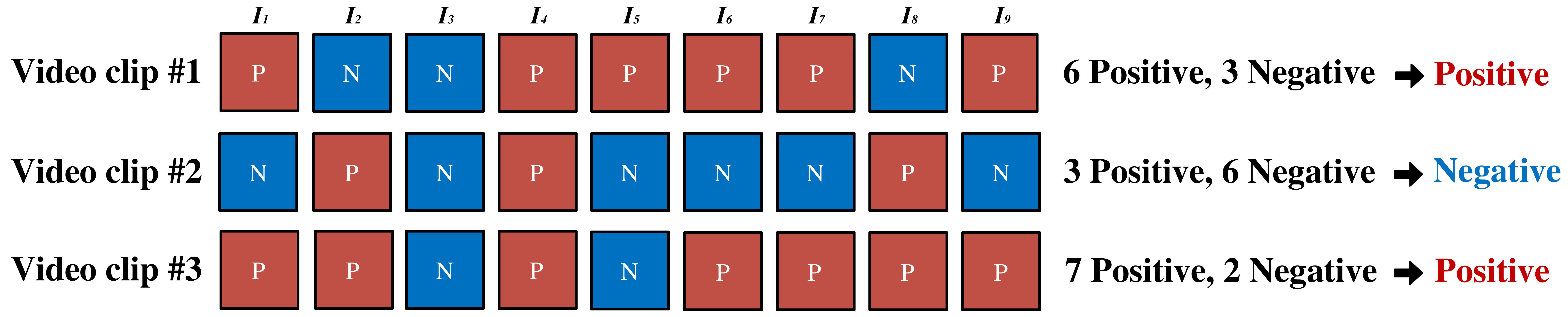}}
\caption{Illustration of the process of majority voting ensemble to improve detection performance.}
\label{fig_majority_voting_2}
\end{figure}

Fig.~\ref{fig_majority_voting_1} is an example of how $I_{n}$ is extracted to detect forged video samples. Because six consecutive frames are required to compose one $I_{n}$ for the training, it is possible to use several $I_{n}$ in one video as inputs of FCDNet for the testing. The final result of FCDNet is summarized using the majority voting ensemble method from each $I_{n}$ extracted in the random position of the video to improve detection performance.
Fig.~\ref{fig_majority_voting_2} is an example of using the majority voting ensemble in practice when nine results of network outputs for one suspicious video are given. While nine $I_{n}$ for one video were put into the trained FCDNet, video clip \#1 had six $I_{n}$ judged as positive and three $I_{n}$ judged as negative. After judging the six positive videos, we finally determine that video clip \#1 is a positive (forged) video with frame-rate modification. In the same way, video clip \#2 is regarded as a negative (original) video with no frame-rate modification because there are six negative and three positive results.

\subsection{Implementation Details}

This section specifies the implementation details of FCDNet, as shown in Fig.~\ref{figure_architecture}. There exist five Block 1's after residual feature generation. We also put five Block 2's to extract low-level information by maintaining the feature dimensions. Moreover, four Block 3's and one Block 4 follow sequentially to extract high-level features from low-level features and classify the FRUC forgery. The number of blocks of each network block was set after the various experiments.
Table~\ref{table_ratio_frame} illustrates the ratio of the original frame and forged frame according to the FRUC method. When the FRUC of the NNI method with 25$\rightarrow$30 fps has occurred, there is at least one interpolated frame between six consecutive frames. For efficiently learning the interpolation traces generated by frame-rate transformation, it is necessary to utilize as many forgery cases as possible for learning. Therefore, we set the number of input frames for the proposed FCDNet to six.

Block 1 has a higher channel capacity ($c$=60) to extract more spatial-domain information in the early stage of FCDNet. On the other hand, Block 2 uses a lower channel capacity ($c$=30) compared with Block 1 to reduce memory requirements and to produce a lightweight FCDNet.
We apply batch normalization \cite{DBLP:journals/corr/IoffeS15} to make FCDNet faster and more stable through the input layer normalization by re-centering and re-scaling.
Moreover, all non-linear activation functions are ReLU \cite{relu}, which overcomes the vanishing gradient problem, allowing the model to learn faster and perform more effectively.
FCDNet has 218,010 parameters and computational complexity of 493.14G MACs (multiply-accumulate operation).

\begin{table}[h]
\renewcommand{\arraystretch}{1.7}
\caption{The ratio of original and forged frames according to the FRUC with various frame interpolation schemes.}
\label{table_ratio_frame}
\centering
\small
\begin{tabular}{|c|c|c|c|c|c|c|}
\hline
\multicolumn{1}{|c|}{\multirow{3}{*}{\shortstack[c]{Frame-rate \\ (fps)}}} & \multicolumn{6}{c|}{Frame interpolation scheme} \\ \cline{2-7} 
& \multicolumn{2}{c|}{NNI} & \multicolumn{2}{c|}{BI} & \multicolumn{2}{c|}{MCI} \\ \cline{2-7} 
 & Original & Forged & Original & Forged & Original & Forged \\ \hline
15 $\rightarrow$ 20 & $\frac{3}{4}$ & $\frac{1}{4}$ & $\frac{1}{4}$ & $\frac{3}{4}$ & $\frac{1}{4}$ & $\frac{3}{4}$ \\
    \cline{1-7} 
15 $\rightarrow$ 25 & $\frac{3}{5}$ & $\frac{2}{5}$ & $\frac{1}{5}$ & $\frac{4}{5}$ & $\frac{1}{5}$ & $\frac{4}{5}$ \\
    \cline{1-7} 
15 $\rightarrow$ 30 & $\frac{1}{2}$ & $\frac{1}{2}$ & $\frac{1}{2}$ & $\frac{1}{2}$ & $\frac{1}{2}$ & $\frac{1}{2}$ \\
    \hline
20 $\rightarrow$ 25 & $\frac{4}{5}$ & $\frac{1}{5}$ & $\frac{1}{5}$ & $\frac{4}{5}$ & $\frac{1}{5}$ & $\frac{4}{5}$ \\
    \cline{1-7} 
20 $\rightarrow$ 30 &  $\frac{2}{3}$ & $\frac{1}{3}$ & $\frac{1}{3}$ & $\frac{2}{3}$ & $\frac{1}{3}$ & $\frac{2}{3}$ \\
    \hline
25 $\rightarrow$ 30 & $\frac{5}{6}$ & $\frac{1}{6}$ & $\frac{1}{6}$ & $\frac{5}{6}$ & $\frac{1}{6}$ & $\frac{5}{6}$ \\
    \hline
\end{tabular}%
\end{table}

\section{Experiments} 
\label{sec_experiments}
In this section, we explain how the dataset is organized for learning and how to evaluate the model's performance. We also introduce the details of how to set up the training environments of the proposed FCDNet.
Finally, a comparison of the prediction results for various FRUC detection methods is presented, demonstrating that FCDNet is the superior method.

\subsection{Settings}
\subsubsection{Datasets}

In the experiments, we used uncompressed videos from the following dataset: XIPH1\textsupsub{$36$}{$352$$\times$$288$}, XIPH2\textsupsub{$22$}{$1920$$\times$$1080$}, and MCL-V\textsupsub{$12$}{$1920$$\times$$1080$}, where the superscripts and subscripts indicate the number of videos and resolution of each database \cite{xiph,mclv}, respectively (Table~\ref{table_uncompressed_cropped_video}). 
The original videos are obtained as cropping raw videos into 256$\times$256 without overlap and encoded by the \textit{libx264} codec of FFMPEG \cite{tomar2006converting}, a free and open-source suite of libraries and programs for handling video, audio, and other multimedia files and streams.
The encoding environment is as follows: group of picture (GOP) is set with 20, B-frame is not used, and the constant rate factor (CRF) value is within \{5, 8, 11, 14\}, where a lower CRF value generally leads to higher quality. Other parameters used the default setting of the FFMPEG.
The frame-rate modified video was created by decoding the original video, interpolating video frames, and re-encoding. 
We convert the video to a specified frame-rate using three types of motion interpolation modes provided by FFMPEG (NNI:`dup', BI:`blend', and MCI:`mci'). 

\begin{table}[h]
	\caption{Details of uncompressed and cropped videos used in train, validation, and test.}
	\centering
	\small
	\begin{tabular} {|c|c|c|c|c|c|c|}
		\hline
		\multicolumn{3}{|c|}{Uncompressed video} & \multicolumn{4}{c|}{Cropped video ($256\times256$)}\\
		\hline
		Dataset & Size & \# of videos & Train & Validation & Test & Total\\
		\hline
		XIPH1 & $352\times288$ & 36 & - & - & 36 & 36\\
		\hline
		XIPH2 & $1080\times1920$ & 22 & 532 & 28 & 56 & 616\\
		\hline
		MCL-V & $1080\times1920$ & 12 & 280 & 28 & 28 & 336\\
		\hline
		\multicolumn{2}{|c|}{Total} & 34 & 812 & 56 & 120 & 988\\
		\hline
	\end{tabular}
	\label{table_uncompressed_cropped_video}
\end{table}

\begin{table}[h]
	\caption{Details of original and frame-rate converted videos used in train, validation, and test.}
	\centering
	\small
	\begin{tabular} {|c|c|c|c|c|c|}
		\hline
		\multirow{6}{*}{Original} & Frame-rate (fps) & CRF value & Train & Validation & Test \\
		\cline{2-6} 
		& 15 & \multirow{4}{*}{\shortstack[c]{5, 8, 11, 14}} & 3,248 & 224 & 480 \\
		\cline{2-2}\cline{4-6}
		& 20 &  & 3,248 & 224 & 480 \\
		\cline{2-2}\cline{4-6}
		& 25 & & 3,248 & 224 & 480 \\
		\cline{2-2}\cline{4-6}
		& 30 &  & 3,248 & 224 & 480 \\
		\cline{2-6}
		& \multicolumn{2}{c|}{Total} & 12,992 & 896 & 1,920 \\
		\hline
		\hline
		\multirow{7}{*}{FRUC} & 15 $\rightarrow$ 20 & \multirow{6}{*}{\shortstack[c]{5, 8, 11, 14}} & 3,248 & 224 & 480 \\
		\cline{2-2}\cline{4-6}
		& 15 $\rightarrow$ 25 &  & 3,248 & 224 & 480 \\
		\cline{2-2}\cline{4-6}
		& 15 $\rightarrow$ 30 &  & 3,248 & 224 & 480 \\
		\cline{2-2}\cline{4-6}
		& 20 $\rightarrow$ 25 &  & 3,248 & 224 & 480 \\
		\cline{2-2}\cline{4-6}
		& 20 $\rightarrow$ 30 &  & 3,248 & 224 & 480 \\
		\cline{2-2}\cline{4-6}
		& 25 $\rightarrow$ 30 &  & 3,248 & 224 & 480 \\
		\cline{2-6}
		& \multicolumn{2}{c|}{Total} & 19,488 & 1,344 & 2,880 \\
		\hline
	\end{tabular}
	\label{table_original_fruc_video}
\end{table}

For an intense focus on frame-rate modification traces, the same encoding environment was set for both generating original video and generating frame-rate converted video by re-encoding after frame interpolation.
The frame-rates of original video and modified video were selected for four cases (15fps, 20fps, 25fps, and 30fps) and six cases (15$\rightarrow$20fps, 15$\rightarrow$25fps, 15$\rightarrow$30fps, 20$\rightarrow$25fps, 20$\rightarrow$30fps, and 25$\rightarrow$30fps) according to the combination, respectively (Table~\ref{table_original_fruc_video}).

\begin{figure}[t]
    \centering%
    \subfigure[]{%
    \includegraphics[width=\linewidth]{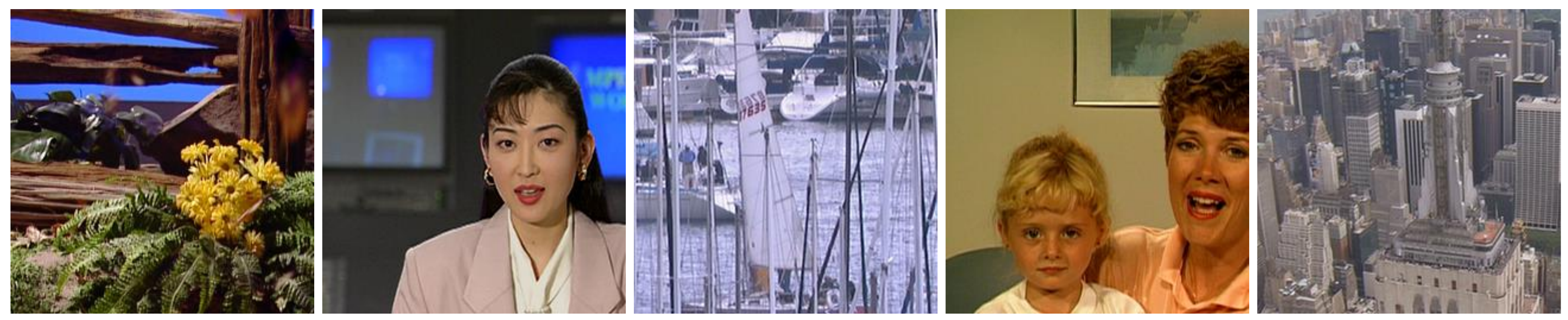}%
    \label{fig_dataset:a}}
    \subfigure[]{%
    \includegraphics[width=\linewidth]{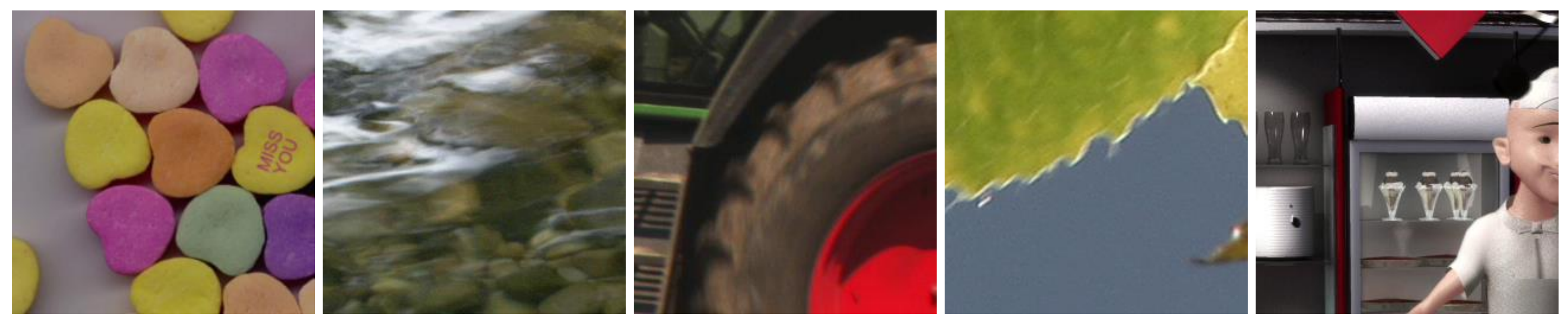}%
    \label{fig_dataset:b}}
    \subfigure[]{%
    \includegraphics[width=\linewidth]{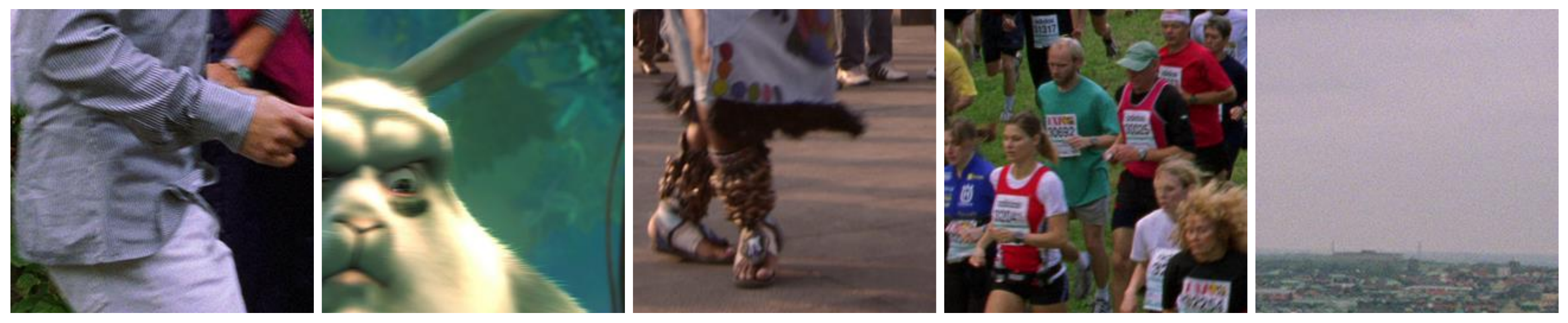}%
    \label{fig_dataset:c}} 
    \caption{Examples of the video contents are from each dataset: (a) XIPH1, (b) XIPH2, and (c) MCL-V. (a) XIPH1 is an non-cropped and unseen dataset for testing. (b) XIPH2 and (c) MCL-V dataset are cropped for training.}
    \label{fig_dataset}
\end{figure}

The original and modified videos were encoded with four CRF parameters in the variable bit rate (VBR) mode of FFMPEG used to obtain a much higher overall quality when file size or average bit rate are not constrained. Consequently, 13,888 (=868$\times$4$\times$4) original videos and 62,496 (=868$\times$4$\times$6$\times$3) forged videos were used in the training process, thus total 76,384 training videos are generated. Similarly, we obtain the following test dataset: 1,920 (=120$\times$4$\times$4) original videos and 8,640 (=120$\times$4$\times$6$\times$3) forged videos were used in the testing process, thus total 10,560 test videos are generated. 
We extracted six successive frames per video randomly for the training and nine groups of clips included six consecutive frames per video for the testing. The ratio of original video and forged video in the training process is 1:1 using a resampling technique such as undersampling of the majority class. Subsequent to the training, the test was conducted using a cropped test dataset (XIPH2\textsupsub{$22$}{$1920$$\times$$1080$} and MCL-V\textsupsub{$12$}{$1920$$\times$$1080$}) and unseen test dataset (XIPH1\textsupsub{$36$}{$352$$\times$$288$}) to evaluate the classification ability and quality of the trained FCDNet prediction.
The examples of the video contents in each dataset are depicted in Fig.~\ref{fig_dataset}.

\captionsetup[figure]{position=top}

\subsubsection{Evaluation Metric}
Detection performance is measured by the true negative rate (TNR), which is the proportion of identified original among the real original videos, the true positive rate (TPR), which is the proportion of identified forged among the real forged videos, and the $F_{1}$ score \cite{manning2008introduction,powers2020evaluation}, which is the harmonic mean of precision and recall. 
Forged frames using FRUC and original frames are denoted as positive and negative, respectively.
The precision is the number of correctly identified positive results divided by the number of all positive results, including those not identified correctly, and the recall is the number of correctly identified positive results divided by the number of all samples that should have been identified as positive. The $F_{1}$ score is usually more useful than accuracy rate, the proportion of observations that have been correctly classified, especially for an uneven class distribution. 
Based on the confusion matrix, the performance metrics were calculated using the formula below.

\begin{equation}
\label{Eq6}
	TNR (\%) = \dfrac{TN}{TN + FP}\times100, \\
\end{equation}
\begin{equation}
\label{Eq7}
	TPR (\%) = \dfrac{TP}{TP + FP}\times100, \\
\end{equation}
\begin{equation}
\label{Eq8}
\begin{aligned}
    F_{1} (\%) = 2\times (\frac{precision \times recall}{precision+recall})\times100, \\[3pt]
    (\ precision = \textstyle \frac{TP}{TP + FP}, \quad \textstyle recall = \frac{TP}{TP + FN}\ ) \\
\end{aligned}
\end{equation}
where $TP$, $TN$, $FP$, and $FN$ are the numbers corresponding to true positives, true negatives, false positives, and false negatives, respectively.
For any classifier, there is always a trade-off between TPR and TNR.
In the FRUC detection case with extremely imbalanced data, the rare class (forged video) is frequently of great interest.
Therefore, we would like to obtain the classification property of FCDNet that produces a high prediction accuracy for TPR, while maintaining reasonable accuracy for TNR.

\begin{sidewaystable}[ph!]
    \vspace*{12cm}
    \caption{Performance evaluation and comparison of input data type and the number of frames in FCDNet with respect to FRUC cases using three frame interpolation schemes.} 
    \centering
    \small
        \begin{tabular}{|c|c|c|c|c|c|c|c|c|c|c|}
        \hline
        \multirow{2}{*}{Input data} & \multirow{2}{*}{\shortstack[c]{Interpolation scheme}} & \multirow{2}{*}{TNR (\%)} & \multicolumn{7}{c|}{TPR (\%)} & \multirow{2}{*}{$F_{1}$ (\%)} \\ \cline{4-10}
         &  &  & 15$\rightarrow$20 & 15$\rightarrow$25 & 15$\rightarrow$30 & 20$\rightarrow$25 & 20$\rightarrow$30 & 25$\rightarrow$30 & Average &  \\ \hline
        \multirow{3}{*}{3 Y-channel} & NNI & \multirow{3}{*}{61.98} & 64.85 & 85.42 & 86.81 & 63.19 & 79.17 & 56.94 & 72.73 & 73.44 \\ \cline{2-2} \cline{4-11} 
         & BI &  & 91.67 & 96.53 & 96.53 & 88.89 & 95.14 & 87.50 & 92.71 & 85.03 \\ \cline{2-2} \cline{4-11} 
         & MCI &  & 88.19 & 91.67 & 93.75 & 84.72 & 88.89 & 81.25 & 88.08 & 82.54 \\ \hline
        \multicolumn{2}{|c|}{Average} & 61.98 & 81.57 & 91.20 & 92.36 & 78.93 & 87.73 & 75.23 & 84.51 & 80.33 \\ \hline\hline
        \multirow{3}{*}{6 Y-channel} & NNI & \multirow{3}{*}{65.10} & 87.50 & 97.22 & 97.22 & 84.72 & 92.36 & 78.47 & 89.58 & 84.18 \\ \cline{2-2} \cline{4-11} 
         & BI &  & 95.83 & 100.00 & 99.31 & 93.75 & 97.22 & 91.67 & 96.30 & 87.22 \\ \cline{2-2} \cline{4-11} 
         & MCI &  & 94.44 & 98.61 & 96.53 & 88.89 & 95.14 & 88.19 & 93.63 & 86.34 \\ \hline
        \multicolumn{2}{|c|}{Average} & 65.10 & 92.59 & 98.61 & 97.68 & 89.12 & 94.91 & 86.11 & 93.17 & 86.08 \\ \hline\hline
        \multirow{3}{*}{2 residual} & NNI & \multirow{3}{*}{73.78} & 73.61 & 97.92 & 100.00 & 65.28 & 92.36 & 49.31 & 79.75 & 80.87 \\ \cline{2-2} \cline{4-11} 
         & BI &  & 83.33 & 92.36 & 86.11 & 83.33 & 85.42 & 79.86 & 85.07 & 84.00 \\ \cline{2-2} \cline{4-11} 
         & MCI &  & 73.61 & 88.89 & 79.86 & 73.61 & 84.03 & 71.53 & 78.59 & 80.16 \\ \hline
        \multicolumn{2}{|c|}{Average} & 73.78 & 76.85 & 93.06 & 88.66 & 74.07 & 87.27 & 66.90 & 81.13 & 81.68 \\ \hline\hline
        \multirow{3}{*}{5 residual} & NNI & \multirow{3}{*}{90.10} & 100.00 & 100.00 & 100.00 & 100.00 & 100.00 & 100.00 & 100.00 & 96.81 \\ \cline{2-2} \cline{4-11} 
         & BI &  & 95.83 & 95.14 & 90.28 & 92.36 & 97.22 & 90.97 & 93.63 & 93.53 \\ \cline{2-2} \cline{4-11} 
         & MCI &  & 91.67 & 95.14 & 81.94 & 90.97 & 96.53 & 86.81 & 90.51 & 91.84 \\ \hline
        \multicolumn{2}{|c|}{Average} & 90.10 & 95.83 & 96.76 & 90.74 & 94.44 & 97.92 & 92.59 & 94.71 & 94.06 \\ \hline
        \end{tabular}%
    \label{table_input_data}
\end{sidewaystable}

\subsubsection{Training Settings}
\label{sec_training_settings}

In the experiments, a data augmentation technique was applied to the training dataset (each original and forged video pair) such as random rotation (e.g., $90^{\circ}$, $180^{\circ}$, and $270^{\circ}$) and horizontal flip.
We improve the learning speed and performance by adopting the paired mini-batch technique, a training strategy in forensics and steganalysis \cite{park2018paired}.
In other words, the pair of original and forged video are selected to construct a single mini-batch in the network training. 
For paired mini-batch training, we select the same frame index number of original and forged videos at the stage of input frames construction.
The size of the paired mini-batch was set to 64 based on the GPU memory size.
FCDNet were trained using the Adam optimizer \cite{adam} with an initial learning rate of $10^{-4}$ and a weight decay of $10^{-5}$ until 50 epochs.
The proposed FCDNet was implemented using the PyTorch framework with two sets of NVIDIA TITAN RTX GPU.

\subsection{Performance Evaluation on Input Data Type}
We should choose the number of input frames and the types of input frames, such as continuous only Y-channel frames or residual Y-channel frames, to increase the detection accuracy of FCDNet. We accomplish this task by analyzing the testing results using a different number of consecutive Y-channel frames and residual Y-channel frames as an input.
Table~\ref{table_input_data} presents the results of accuracy as to the various types of input features on FCDNet.
The use of five residual Y-channel frames as one input stack has higher accuracy than other input data types.

\subsection{Performance Evaluation on Majority Voting Ensemble}
Table~\ref{table_majority_voting} illustrates the test accuracy according to the number of $I_{n}$ used for the frame-rate forgery determination. 
During testing, we use the majority voting principle method to combine the predictions from multiple other outputs. Majority voting ensemble is a technique that might be used to improve the performance, ideally achieving higher performance than any single output. It involves summing the predictions for each class label and predicts the class label with the most votes.
This majority voting ensemble method does not require additional training of FCDNet, but provides additional performance gain.
The $F_{1}$ was 93.78\% using only one $I_{n}$, and the accuracy increased by 2.38\% trying to enlarge the number of $I_{n}$.
When the number of $I_{n}$ is greater than nine, the detection accuracy converges to some extent.
But, as the computation time per video continues to increase, we experimentally set the number of $I_{n}$ for a majority voting ensemble to nine.

\begin{table}[h]
\caption{Effects of majority voting ensemble. We report the prediction result of FCDNet with respect to the number of input data $I_{n}$. Parentheses indicate the ratio of computation time per video between baseline and each the number of $I_{n}$.}
    \centering
    \small
    \begin{tabular} {|c|c|c|c|c|}
    	\hline
    	\# of $I_{n}$ & TNR (\%) & TPR (\%) & $F_{1}$ (\%) & Computation time per video (s)\\
    	\hline
    	1 & 84.55 & 91.32 & 93.78 & 0.018 (baseline)\\
    	\hline
    	5 & 89.24 & 94.14 & 95.80 & 0.057 ($\times$3.17) \\
    	\hline
    	7 & 88.54 & 94.17 & 95.74 & 0.077 ($\times$4.28)\\
    	\hline
    	9 & 90.10 & 94.64 & 96.16 & 0.097 ($\times$5.39) \\
    	\hline
    	11 & 89.93 & 94.14 & 96.12 & 0.114 ($\times$6.33)\\
    	\hline
    	13 & 89.58 & 94.56 & 96.06 & 0.134 ($\times$7.44)\\
    	\hline
    \end{tabular}
    \label{table_majority_voting}
\end{table}

\begin{sidewaystable}[ph!]
    \vspace*{12cm}
    \caption{Performance evaluation of FCDNet and previous methods for frame interpolation schemes with respect to FRUC cases using three frame interpolation schemes.}
    \centering
    \small
        \begin{tabular}{|c|c|c|c|c|c|c|c|c|c|c|}
        \hline
        \multirow{2}{*}{Method} & \multirow{2}{*}{\shortstack[c]{Interpolation scheme}} & \multirow{2}{*}{TNR (\%)} & \multicolumn{7}{c|}{TPR (\%)} & \multirow{2}{*}{$F_{1}$ (\%)} \\ \cline{4-10}
         &  &  & 15$\rightarrow$20 & 15$\rightarrow$25 & 15$\rightarrow$30 & 20$\rightarrow$25 & 20$\rightarrow$30 & 25$\rightarrow$30 & Average &  \\ \hline
        \multirow{3}{*}{Bestagini \cite{Bestagini}} & NNI & \multirow{3}{*}{90.09} & 91.38 & 97.41 & 94.83 & 87.07 & 92.24 & 75.86 & 89.80 & 89.93 \\ \cline{2-2} \cline{4-11} 
         & BI &  & 37.93 & 73.28 & 43.97 & 39.66 & 55.17 & 18.97 & 44.83 & 57.94 \\ \cline{2-2} \cline{4-11} 
         & MCI &  & 36.21 & 75.00 & 39.66 & 50.86 & 59.48 & 19.83 & 46.84 & 59.76 \\ \hline
        \multicolumn{2}{|c|}{Average} & 90.09 & 55.17 & 81.90 & 59.48 & 59.20 & 68.97 & 38.22 & 60.49 & 69.21 \\ \hline\hline
        \multirow{3}{*}{Jung \cite{Jung}} & NNI & \multirow{3}{*}{90.09} & 100.00 & 100.00 & 100.00 & 100.00 & 100.00 & 100.00 & 100.00 & 95.28 \\ 
        \cline{2-2} \cline{4-11} 
         & BI &  & 93.10 & 96.55 & 83.62 & 71.55 & 92.24 & 60.34 & 82.90 & 85.99 \\ 
         \cline{2-2} \cline{4-11} 
         & MCI &  & 93.10 & 100.00 & 75.86 & 86.21 & 100.00 & 70.69 & 87.64 & 88.73 \\ \hline
        \multicolumn{2}{|c|}{Average} & 90.09 & 95.40 & 98.85 & 86.49 & 85.92 & 97.41 & 77.01 & 90.18 & 90.00 \\ \hline\hline
        \multirow{3}{*}{FCDNet} & NNI & \multirow{3}{*}{90.10} & 100.00 & 100.00 & 100.00 & 100.00 & 100.00 & 100.00 & 100.00 & 96.81 \\
        \cline{2-2} \cline{4-11} 
         & BI &  & 95.83 & 95.14 & 90.28 & 92.36 & 97.22 & 90.97 & 93.63 & 93.53 \\ \cline{2-2} \cline{4-11} 
         & MCI &  & 91.67 & 95.14 & 81.94 & 90.97 & 96.53 & 86.81 & 90.51 & 91.84 \\ \hline
        \multicolumn{2}{|c|}{Average} & 90.10 & 95.83 & 96.76 & 90.74 & 94.44 & 97.92 & 92.59 & 94.71 & 94.06 \\ \hline
        \end{tabular}%
    \label{table_perfom_comparison_existing_method}
\end{sidewaystable}

\subsection{Performance Evaluation Compared with Conventional Method}

The comparative experiment results of FRUC detection are depicted in Table~\ref{table_perfom_comparison_existing_method}.
We use the MATLAB source code of Jung \textit{et al.} and the experiment settings of the comparative hand-designed and feature-based approaches were set as described in each paper \cite{Bestagini,Jung}. 
The previous methods use a key threshold to measure the performance that must be determined. 
Because the performance of TNR and TPR varies depending on the key threshold value, it is important to set an appropriate threshold to apply in real-world applications. We set the thresholds to 3.359 and 3.832 for the Bestagini method and Jung method, respectively. 
These schemes judge the original video as pristine at 90.09\% that is familiar with the TNR of the proposed FCDNet.
Both NNI and MCI cases demonstrated similar performance compared with the Jung method, but BI case demonstrated significantly improved performance in FCDNet. Regarding the results depicted in Table~\ref{table_perfom_comparison_existing_method}, both the Jung method and the proposed method illustrate significantly higher performance than the Bestagini method. Compared with the Jung method with a similar TNR of 90.10\%, there was a 4.53\% increase in the average TPR and also a 4.06\% point increase in the average $F_{1}$ score.

The comparison of the video frame-rate conversion detection results according to the type of test dataset are depicted in Table~\ref{table_test_data}. We cropped a large-resolution video to 256$\times$256, and then the cropped video was used to train FCDNet. In addition, the test was separately conducted on both the cropped video dataset (i.e., XIPH2 and MCL-V) and the non-cropped video dataset (i.e., XIPH1) not used in the learning process. The previous methods, those of Jung and Bestagini, exhibited high performance for the entire video without being cropped, but the performance was significantly degraded in the cropped video. However, the proposed FCDNet had high $F_{1}$ scores over 94.06\% and 98.22\% for both test datasets.
It can be seen that FCDNet accurately learned the pattern occurring in the spatial and temporal domain due to the artifacts of the frame interpolation scheme. Furthermore, FCDNet does not require content information such as the object and the background or foreground in the entire video. We can judge the integrity using not all of video but just a cropped video based on FCDNet. This capability is a significant advantage in a real environment because FCDNet can derive the video forensic results from 
suspicious contents regardless of input video size.

\begin{table}[h]
\renewcommand{\arraystretch}{1.3}
	\caption{Prediction result comparison of FCDNet and previous methods with respect to various test dataset types: non-cropped dataset vs. cropped dataset.}
	\centering
	\small
	\begin{tabular} {|c|c|c|c|c|c|c|}
		\hline
		\multirow{2}{*}{Method} & \multicolumn{3}{c|}{Non-cropped dataset} & \multicolumn{3}{c|}{Cropped dataset}\\
		\cline{2-7}
		& TNR (\%) & TPR (\%) & $F_{1}$ (\%) & TNR (\%) & TPR (\%) & $F_{1}$ (\%)  \\
		\hline
		Bestagini \cite{Bestagini} & 90.09 & 60.49 & 69.21 & 90.51 & 40.11 & 52.16\\
		\hline
	    Jung \cite{Jung}  & 90.09 & 90.18 & 90.00 & 92.47 & 64.88 & 72.84\\
		\hline
		FCDNet & 90.10 & 94.71 & 94.06 & 99.38 & 96.64 & 98.22\\
		\hline
	\end{tabular}
	\label{table_test_data}
\end{table}

The comparison of average processing time between FCDNet and the previous methods is summarized in Table~\ref{table_detection_speed}. 
We check the speed by measuring the time required for FRUC detection, excluding the pre-processing time.
The Bestagini and Jung methods use all the video frames, whereas the proposed FCDNet only requires 54 frames (= 9$\times$6) as a bundled input. Consequently, whereas the Bestagini and Jung methods require at least 100 frames to obtain acceptable detection performance, the proposed FCDNet can achieve higher detection accuracy using a small number of frames.
FCDNet can determine whether a suspicious video is forged with high accuracy and have a short process time using only nine stacks of successive residual frames.
The Jung method requires more than 11.737 seconds per video clip, but the proposed FCDNet requires approximately 0.097 seconds. Therefore, it can be confirmed that FCDNet is about 65.34 times faster than the Jung method.

\begin{table}[h]
\renewcommand{\arraystretch}{1.3}
	\caption{Evaluation of computation time of FCDNet and previous methods.}
	\centering
    \small
	\begin{tabular} {|c|c|c|c|}
		\hline
		Evaluation criteria & Bestagini \cite{Bestagini}& Jung \cite{Jung}& FCDNet \\
		\hline
		Min. \# of frames & 100 & 100 & 54\\
		\hline
		Computation time per frame (s) & 0.095 & 0.117 & 0.002 \\
		\hline
		Computation time per video (s) & 9.490 & 11.737 & 0.097\\
		\hline
	\end{tabular}
	\label{table_detection_speed}
\end{table}

\subsection{Performance Evaluation Compared with Comparative Networks}
 
For presenting the effectiveness of the FCDNet architecture, we designed an experiment to analyze the classification performance of our approach compared with HeNet \cite{henet} suggested for double compression detection and MesoNet \cite{afchar2018mesonet} proposed for deepfake detection. As there was no video FRUC detection research based on neural network, we selected other CNN-based networks to target video tempering as a comparative network. 
The network structures and hyperparameters of those comparative networks were set as described in each paper. The CNN components were lightly modified because of the input data, a stack of five consecutive residual frames, to utilize the identical input parameter with FCDNet.
For fair experiments, the data augmentation, the weight initialization, the batch size, and optimizer were set equally using the methodology specified in Section~\ref{sec_training_settings}.

\begin{sidewaystable}
    \vspace*{12cm}
    \caption{Performance evaluation of the proposed FCDNet and comparative CNNs for frame interpolation schemes with respect to FRUC cases.}
    \centering
    \small
        \begin{tabular}{|c|c|c|c|c|c|c|c|c|c|c|}
        \hline
        \multirow{2}{*}{Method} & \multirow{2}{*}{\shortstack[c]{Interpolation scheme}} & \multirow{2}{*}{TNR (\%)} & \multicolumn{7}{c|}{TPR (\%)} & \multirow{2}{*}{$F_{1}$ (\%)} \\ \cline{4-10}
         &  &  & 15$\rightarrow$20 & 15$\rightarrow$25 & 15$\rightarrow$30 & 20$\rightarrow$25 & 20$\rightarrow$30 & 25$\rightarrow$30 & Average &  \\ \hline
        \multirow{3}{*}{HeNet \cite{henet}} & NNI & \multirow{3}{*}{50.69} & 80.56 & 83.33 & 88.19 & 79.86 & 86.81 & 73.61 & 82.06 & 76.36 \\ \cline{2-2} \cline{4-11} 
         & BI &  & 95.83 & 98.61 & 98.61 & 95.14 & 96.53 & 95.14 & 96.64 & 84.22 \\ \cline{2-2} \cline{4-11} 
         & MCI &  & 86.11 & 93.75 & 95.83 & 87.50 & 91.67 & 85.42 & 90.05 & 80.79 \\ \hline
        \multicolumn{2}{|c|}{Average} & 50.69 & 87.50 & 91.90 & 94.21 & 87.50 & 91.67 & 84.72 & 89.58 & 80.45 \\ \hline\hline
        \multirow{3}{*}{MesoNet \cite{afchar2018mesonet}} & NNI & \multirow{3}{*}{62.33} & 81.25 & 86.81 & 88.19 & 75.00 & 81.94 & 69.44 & 80.44 & 78.27 \\ \cline{2-2} \cline{4-11} 
         & BI &  & 95.14 & 98.61 & 96.53 & 94.44 & 96.53 & 93.75 & 95.83 & 86.75 \\ \cline{2-2} \cline{4-11} 
         & MCI &  & 88.89 & 97.22 & 90.28 & 87.50 & 93.75 & 88.19 & 90.97 & 84.20 \\ \hline
        \multicolumn{2}{|c|}{Average} & 62.33 & 88.43 & 94.21 & 91.67 & 85.65 & 90.74 & 83.80 & 89.08 & 83.07 \\ \hline\hline
        \multirow{3}{*}{FCDNet} & NNI & \multirow{3}{*}{90.10} & 100.00 & 100.00 & 100.00 & 100.00 & 100.00 & 100.00 & 100.00 & 96.81 \\ \cline{2-2} \cline{4-11} 
         & BI &  & 95.83 & 95.14 & 90.28 & 92.36 & 97.22 & 90.97 & 93.63 & 93.53 \\ \cline{2-2} \cline{4-11} 
         & MCI &  & 91.67 & 95.14 & 81.94 & 90.97 & 96.53 & 86.81 & 90.51 & 91.84 \\ \hline
        \multicolumn{2}{|c|}{Average} & 90.10 & 95.83 & 96.76 & 90.74 & 94.44 & 97.92 & 92.59 & 94.71 & 94.06 \\ \hline
        \end{tabular}%
    \label{table_perfom_comparison_comparative_method}
\end{sidewaystable}

Table~\ref{table_perfom_comparison_comparative_method} present the comparison results using the three frame interpolation schemes for FRUC. 
The average $F_{1}$ score for FCDNet is 94.06\%, which is a 10.99\% higher than MesoNet, which has the second-best performance: 83.07\% average $F_{1}$ score. 
HeNet, which is specialized for video double MPEG-4 compression detection, demonstrated lower performance for detecting subtle traces of video frame-rate conversion. 
Also, MesoNet demonstrated high performance in facial forgery detection but revealed limitations when learning artifacts caused by video frame-rate converting.
The comparative methods have relatively high TPR for BI and MCI because of unique characteristics such as blurring and ghosting artifacts that occurred while interpolating a video frame. However, they have low TPR in NNI case and do not produce satisfactory TNR results compared with FCDNet. Therefore, we confirm that comparative studies have restrictions.
In contrast, FCDNet is more robust than comparative networks in every FRUC case, as listed in Table~\ref{table_perfom_comparison_comparative_method}.

\subsection{Robustness on Unseen FRUC}

The following experiments were conducted to evaluate the robustness against the unlearned frame-rate. Since we do not know the actual frame-rate in real-world environments, it must be able to detect on the unlearned frame-rate parameter. Therefore, we create a dataset for frame-rate conversions of 15$\rightarrow$24, 20$\rightarrow$24, and 24$\rightarrow$30 fps using the same method as training dataset generations. Based on Table~\ref{table_robustness_FR}, for unlearned frame-rate transformation, its $F_{1}$ score decreased by 0.29\% but its accuracy was still high at 93.77\%. FCDNet learned a pattern that appears as a frame-rate transformation rather than a frame-rate parameter. Consequently, FCDNet is robust for the frame-rate parameter.

\begin{table}[t]
	\caption{Prediction result comparison with various frame-rate in FCDNet: learned frame-rate vs. unseen frame-rate. Parentheses indicate difference values between learned and unlearned.}
	\centering
	\small
	\begin{tabular} {|c|c|c|c|c|}
		\hline
		\multicolumn{2}{|c|}{FRUC (fps)} & TNR (\%) & TPR (\%) & $F_{1}$ (\%) \\
		\hline
		\multirow{6}{*}{Learned} & 15 $\rightarrow$ 20 & \multirow{6}{*}{90.10} & \multirow{6}{*}{94.71} & \multirow{6}{*}{94.06}\\
		\cline{2-2}
		& 15 $\rightarrow$ 25 &  &  &  \\
		\cline{2-2}
		& 15 $\rightarrow$ 30 &  &  &  \\
		\cline{2-2}
		& 20 $\rightarrow$ 25 &  &  &  \\
		\cline{2-2}
		& 20 $\rightarrow$ 30 &  &  &  \\
		\cline{2-2}
		& 25 $\rightarrow$ 30 &  &  &  \\
		\hline\hline
		\multirow{3}{*}{Unseen} & 15 $\rightarrow$ 24 & 89.93 (-0.17) & 94.91 (+0.20) & 94.14 (+0.08) \\
		\cline{2-5}
		& 20 $\rightarrow$ 24 & 88.54 (-1.56) & 93.06 (-1.65) & 92.73 (-1.33) \\
		\cline{2-5}
	    & 24 $\rightarrow$ 30 & 88.89 (-1.21) & 94.44 (-0.27) & 94.44 (+0.38) \\
		\hline
		\multicolumn{2}{|c|}{Average} & 89.12 (-0.98) & 94.14 (-0.57) & 93.77 (-0.29) \\ \hline
	\end{tabular}
	\label{table_robustness_FR}
\end{table}

\subsection{Robustness on Unseen CRF Value}

The following experiment was designed to evaluate robustness against the unlearned CRF values. When recompression occurs after the video frame-rate modification, it can be compressed with various CRF values using FFMPEG's VBR mode in the real-world, so we should be able to detect frame-rate modified videos forged with unlearned CRF parameters. In particular, the higher CRF value is used the more difficult to find frame-rate change traces because video compression loss increases as the CRF value increases (strong compression) used for recompression.

We verified and compared this well-known tendency by creating a dataset for frame-rate conversion of the higher CRF value of 17 and the lower CRF value of 2, similar to learned dataset generations. Table~\ref{table_robustness_CRFvalue} reveals that the $F_{1}$ score even improves by 1.60\% in the lower quality video (CRF=17), where a 1.65\% lower TPR but a 2.11\% higher TNR are achieved, respectively. 
Also, it was expected that the higher quality video (CRF=2) improved by 5.32\% for the $F_{1}$ score.
Based on the results for the unseen CRF values, FCDNet demonstrated stable and outstanding performance.

\begin{table}[h]
	\caption{Prediction result comparison with various quality factors in FCDNet: learned CRF value vs. unlearned CRF value. Parentheses indicate difference values between learned and unlearned.}
	\centering
	\small
	\begin{tabular} {|c|c|c|c|c|}
		\hline
		\multicolumn{2}{|c|}{CRF value} & TNR (\%) & TPR (\%) & $F_{1}$ (\%) \\
		\hline
		\multirow{4}{*}{Learned} & 5 & \multirow{4}{*}{90.10} & \multirow{4}{*}{94.71} & \multirow{4}{*}{94.06} \\
		\cline{2-2}
		& 8 &  &  &  \\
		\cline{2-2}
		& 11 &  &  &  \\
		\cline{2-2}
		& 14 &  &  &  \\
		\hline\hline
		\multirow{2}{*}{Unlearned} & 2 & 99.46 (+9.36) & 98.89 (+4.18) & 99.38 (+5.32) \\
		\cline{2-5}
		& 17 & 92.21 (+2.11) & 93.06 (-1.65) & 95.66 (+1.60) \\
		\hline
		\multicolumn{2}{|c|}{Average} & 95.84 (+5.74) & 95.97 (+1.26) & 97.52 (+3.46) \\ \hline
	\end{tabular}
	\label{table_robustness_CRFvalue}
\end{table}

\begin{table}[t]
	\caption{Prediction result comparison with various bit rates of FFMPEG's CBR mode in FCDNet: learned CRF value vs. unlearned CBR mode parameter. Parentheses indicate difference values between learned and unlearned.}
	\centering
	\small
	\begin{tabular} {|c|c|c|c|c|}
		\hline
		\multicolumn{2}{|c|}{Encoding factor} & TNR (\%) & TPR (\%) & $F_{1}$ (\%) \\
		\hline
		\multirow{4}{*}{\shortstack[c]{Learned \\ VBR mode param. \\(CRF values)}} & 5 & \multirow{4}{*}{90.10} & \multirow{4}{*}{94.71} & \multirow{4}{*}{94.06} \\
		\cline{2-2}
		& 8 &  &  &  \\
		\cline{2-2}
		& 11 &  &  &  \\
		\cline{2-2}
		& 14 &  &  &  \\
		\hline\hline
		\multirow{4}{*}{\shortstack[c]{Unlearned \\ CBR mode param. \\(bit rate values)}} & 500k & 68.75 (-21.35) & 90.12 (-4.59) & 91.46 (-2.60) \\
		\cline{2-5}
		& 600k & 70.83 (-19.27) & 90.43 (-4.28) & 92.43 (-1.63) \\
		\cline{2-5}
		& 700k & 72.62 (-17.48) & 91.51 (-3.20) & 92.66 (-1.40) \\
		\cline{2-5}
		& 800k & 76.39 (-13.71) & 91.82 (-2.89) & 93.19 (-0.87) \\
		\hline
		\multicolumn{2}{|c|}{Average} & 72.15 (-17.95) & 90.97 (-3.74) & 92.43 (-1.63) \\ \hline
	\end{tabular}
	\label{table_robustness_CBR}
\end{table}

\subsection{Robustness on Unseen Encoding Factor}

We conducted additional experiments for unlearned encoding factors of video rate control. Rate control decides how many bits will be used for each frame, which determines the file size and also how quality is distributed. In the training phase, we consider the dataset generated by VBR settings, which maintains the best quality and care less about the file size. However, there is a scenario in which the bit rate stays constant over the entire video stream using the constant bit rate (CBR) mode of FFMPEG. For analyzing the robustness of the unseen video encoding mode of FCDNet that learned with various CRF values using VBR mode of FFMPEG, we additionally create a dataset with various bit rate values (i.e., 500k, 600k, 700k, and 800k) using CBR mode in FFMPEG.
The average values of TNR, TPR, and $F_{1}$ scores were 72.15\%, 90.97\%, and 92.43\% in unlearned CBR mode's bit rate values, respectively, as depicted in Table~\ref{table_robustness_CBR}. Compared with the results in VBR mode's CRF values, the detection performance of FCDNet is relatively low. We estimated that this performance degradation was caused by the visual quality deterioration in CBR mode encoding. During VBR encoding, the video file's bit rate will dynamically increase or decrease depending on its needs. The quality of the learned dataset generated by VBR mode is superior compared with the unlearned dataset from CBR mode. Therefore, the performance of FCDNet may deteriorate in the dataset created from the bit rate values of CBR mode encoding.

\begin{figure}[t]%
\centering%
\subfigure[]{%
  \includegraphics[width=0.32\linewidth]{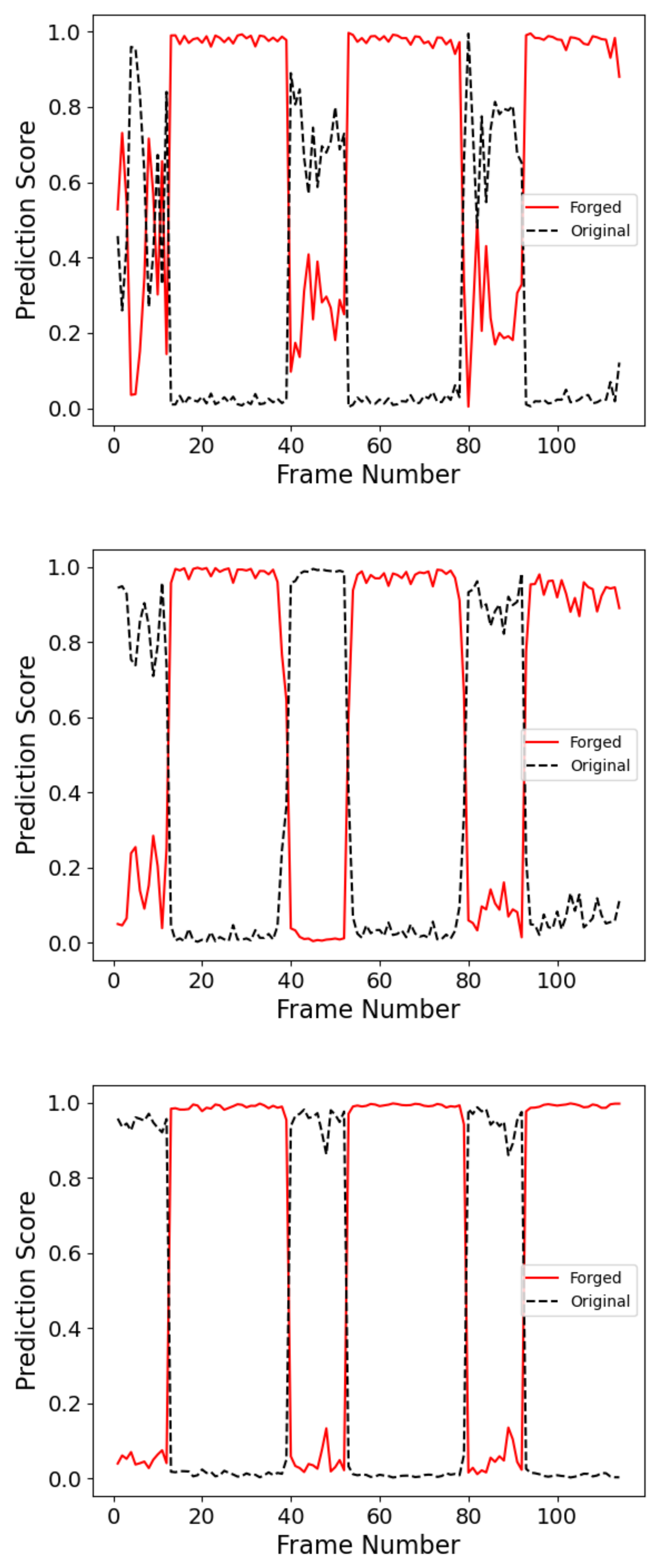}%
  \label{fig_tl:ex:a}
}\hspace{-1mm}
\subfigure[]{%
  \includegraphics[width=0.32\linewidth]{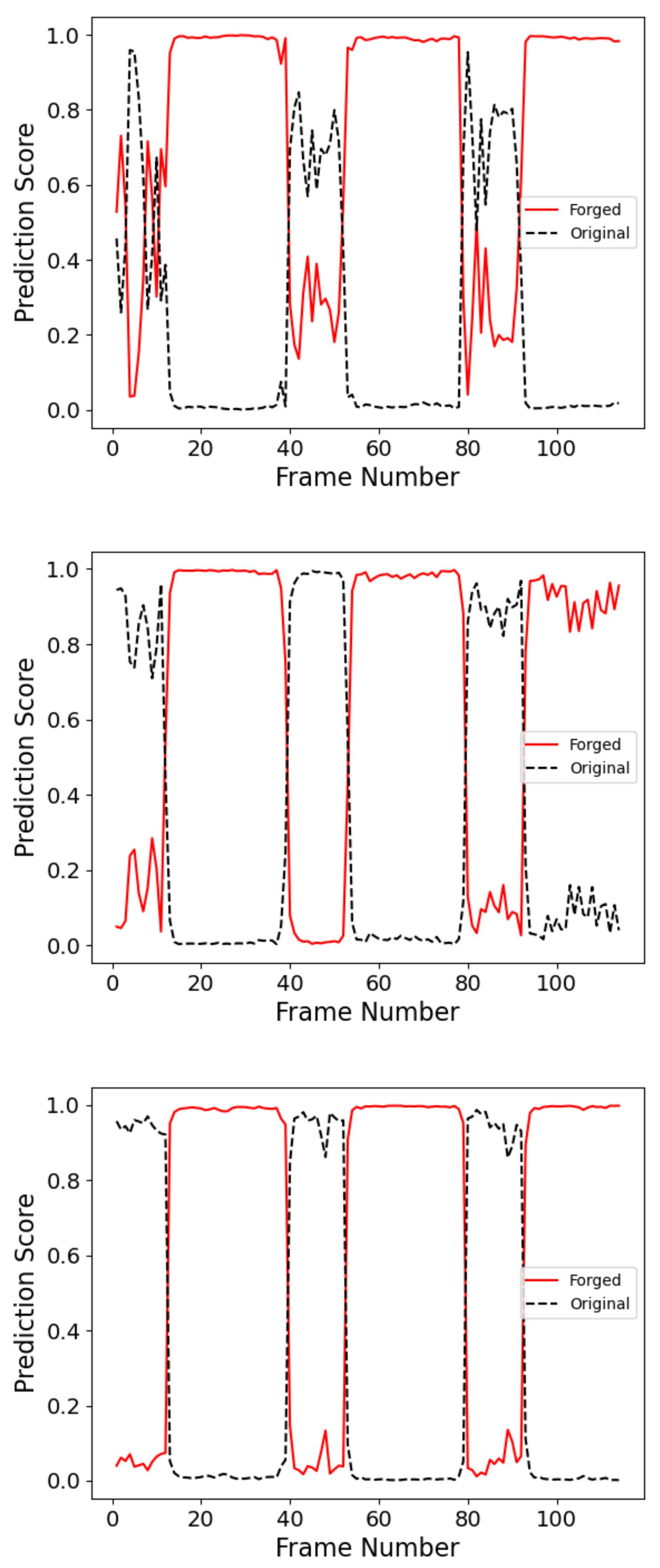}%
  \label{fig_tl:ex:b}
}\hspace{-1mm}
\subfigure[]{%
  \includegraphics[width=0.32\linewidth]{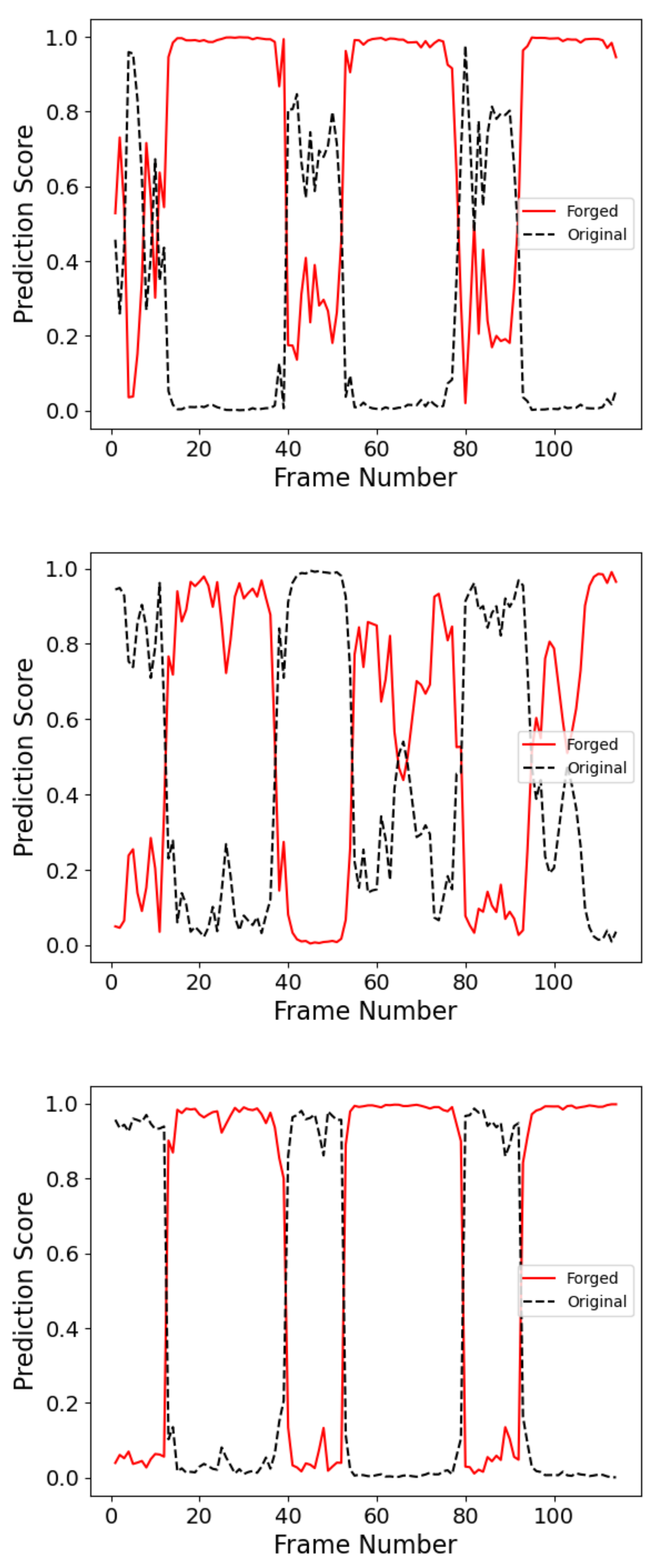}%
  \label{fig_tl:ex:c}
}
\caption{Results of temporal-localization. (a) the result of NNI-based FRUC applied to video, (b) the result of BI-based FRUC applied to video, and (c) the result of MCI-based FRUC applied to video.}
\label{fig_tl}%
\end{figure}

\subsection{Temporal Localization}
This section presents the temporal localization results for frame-rate converted artifacts of FCDNet. For the temporal-localization test, we cropped an original video in the XIPH1\textsupsub{$36$}{$352$$\times$$288$} dataset with 15fps to 256$\times$256 and manipulated the video from 1 to 2 seconds (from the 16-$th$ frame to the 40-$th$ frame), from 3 to 4 seconds (from the 56-$th$ frame to the 80-$th$ frame), and from 5 to 6 seconds (from the 96-$th$ frame to the 120-$th$ frame) with 15$\rightarrow$25 fps, and then recompressed this by the same parameters with first-compression.
With the dataset partially deformed on the time axis as input data, the test process is performed on all video frames using a sliding window size of six with a stride of one.
Fig.~\ref{fig_tl} illustrates the result of temporal localization; the x-axis and the y-axis represent the index of frames and the prediction scores for each class, respectively. 
FCDNet classified forged frames as manipulated and original frames as integrity with considerable accuracy in the time-domain. Therefore, to localize the frame-rate converted frames temporally is possible using FCDNet.

\section{Conclusion}
\label{sec_conclusion}

In this paper, we proposed the end-to-end trainable FCDNet for identifying FRUC, which exploits spatial and temporal information successively. It uses a stack of consecutive residual frames as an input for the video frame-rate conversion detection with higher performance. The proposed FCDNet consists of four types of network blocks, specialized for learning FRUC forensic features. Furthermore, the majority voting ensemble technique for enhancing detection performance was presented. We demonstrated the effectiveness of FCDNet by conducting extensive experiments with various comparative approaches. 
It has the advantage of a higher forgery detection speed because our work can verify integrity accurately with a little part of the video.
Furthermore, FCDNet has robustness on unseen video encoding parameters so that it is suitable for practical forensics in real-world environments.
Additionally, FCDNet can be applied to localize the temporally FRUC position.
In future research, we will search for optimal hyperparameters, such as the number of blocks and channels with a constrained resource, and extend research on frame-rate down conversion cases.

\begin{acknowledgements}
This work was supported by Institute of Information \& communications Technology Planning \& Evaluation (IITP) grant funded by the Ministry of Science and ICT (MSIT) of Korea Government (No. 2017-0-01671, Development of high reliability image and video authentication service for smart media environment).
\end{acknowledgements}

%
%



\end{document}